\documentclass[aps, showpacs, preprintnumbers, prd, nofootinbib, preprint]{revtex4}
\usepackage{feynmf}
\usepackage{amsmath}
\usepackage{graphicx}
\usepackage{mathrsfs}
\usepackage{stmaryrd}
\usepackage{amsfonts}
\usepackage{wrapfig}
\usepackage{subfig}
\usepackage{float}
\usepackage{bbm}

\begin{document}
\def\sech{\mathop{\rm{sech}}\nolimits}
\def\arcsinh{\mathop{\rm{arcsinh}}\nolimits}

\title{Fermion masses and mixing in a 4+1-dimensional $SU(5)$ domain-wall brane model}
\author{Benjamin D. Callen}\email{b.callen@pgrad.unimelb.edu.au}
\affiliation{School of Physics, The University of Melbourne, Victoria 3010, Australia}
\author{Raymond R. Volkas}\email{raymondv@unimelb.edu.au}
\affiliation{School of Physics, The University of Melbourne, Victoria 3010, Australia}

\begin{abstract}

 We study the fermion mass and mixing hierarchy problems within the context of the $SU(5)$ 4+1d domain-wall brane model of Davies, George and Volkas. In this model, the ordinary fermion mass relations of $SU(5)$ grand unified theories are avoided since the masses are proportional to overlap integrals of the profiles of the electroweak Higgs and the chiral components of each fermion, which are split into different 3+1d hyperplanes according to their hypercharges. We show that the fermion mass hierarchy without electroweak mixing can be generated naturally from these splittings, that generation of the CKM matrix looks promising, and that the Cabibbo angle along with the mass hierarchy can be generated for the case of Majorana neutrinos from a more modest hierarchy of parameters. We also show that under some assumptions made on the parameter space, the generation of realistic lepton mixing angles is not possible without fine-tuning, which argues for a flavour symmetry to enforce the required relations.

\end{abstract}

\maketitle

\newpage

\section{Introduction}

 In the standard model (SM), three of the most open problems are how the fermion mass hierarchy is generated, the origin of small mixing angles in the Cabibbo-Kobayashi-Maskawa (CKM) matrix and near tribimaximal mixing in the lepton sector. With neutrino masses now known to be nonzero but under $1\;eV$, the mass hierarchy has a spread of at least 14 orders of magnitude, given that the top quark has a mass of roughly $170 \;GeV$. Amongst approaches used for solving these problems are grand unified theories (GUTs), higher dimensional operators, and flavor symmetries.

 One of the most promising new theoretical frameworks for solving hierarchy problems that has emerged over the last decade has been extra-dimensional models, such as the Arkani--Hamed-Dimopoulos-Dvali (ADD) model \cite{originalbranepaper}, and the two Randall-Sundrum (RS) models \cite{randallsundrum2, randallsundrum1}. The ADD and RS1 frameworks solve the hierarchy problem between the Planck scale and the electroweak scale, which is of a similar order of magnitude to that of the fermion mass spectra. For other papers on extra-dimensional models, see Refs. \cite{antoniadisnewdimattev, newdimatmillimeter, exotickkmodels, gibbonswiltshire, pregeometryakama}.
 
 In RS2 models the gauge hierarchy problem is not solved by extra-dimensional physics, but the split fermion idea of Arkani-Hamed and Schmaltz \cite{splitfermions} can be used to generate fermion mass hierarchies from exponentially sensitive overlap integrals of extra-dimensional profile functions.  Similarly, the RS1 setup can address this problem by allowing fermions to propagate in the bulk and thus acquire non-trivial profiles \cite{grossmanneubertbulkfermions, gherghettapomarol2000}.  The idea is that the 3+1d fermion zero modes are in general localized around different locations along the extra dimension, with dimensional reduction then producing an effective 3+1d Yukawa coupling constant that is the product of the 4+1d Yukawa coupling constant and an overlap integral involving profile functions.  When the profiles are split, the overlap integrals are suppressed, leading to small 3+1d effective Yukawa coupling constants.  This fits in well with the fact that quark and lepton masses, except for the top quark, are {\it suppressed} with respect to the electroweak scale.  Scalar bosons will also in general be split, a phenomenon we shall use to suppress colored-Higgs-induced proton decay (see Refs. \cite{mckellarpdecsuppaper, twistedsplitfermions, cpandtwistedsf, realisticsplitfermions, higgslocinsfmodels, cpin5dchangng, nnbaroscinlargeed, coulombicintbwsplitfermions, kakizakidoublettriplet, kakizakimassfitting} for more on the use of the splitting of  fermions and bosons in extra dimensions to generate fermion mass textures and to suppress proton decay and other baryon number violating processes).

 In this paper, we shall utilize the $SU(5)$ 4+1d domain-wall brane model devised by Davies, George and Volkas (DGV) \cite{firstpaper} to address the fermion mass and mixing angle problems. In this RS2-like model, the split fermion idea arises naturally, and thus the usual $SU(5)$ quark-lepton mass relations are avoided. It will be shown that the mass hierarchy problem can be solved using this method, and that the mass hierarchy and the Cabibbo angle can be accounted for in the two-generation case with Majorana neutrinos. We also explain why tribimaximal mixing cannot be accounted for without fine-tuning, and that the addition of a flavor symmetry therefore seems necessary.  We are thus led to the view that extra dimensions provide an excellent way to qualitatively understand mass hierarchies, but they are insufficient to explain all the observed mixing angle patterns.  The reason the flavor problem has proven to be so difficult may be because more than one ingredient is necessary: extra dimensions on their own, and flavor symmetry on its own, are only partially successful.
 
 The following section reviews the DGV model and develops it further in several important ways: neutrino mass generation is examined and the see-saw mechanism implemented, and the dynamics of scalar-field localization is shown to be analytically tractable.  Section \ref{sec:hierarchy} then analyses the parameter space of the model to produce the required mass and mixing angle hierarchies, with the aforementioned caveat for tribimaximal lepton mixing.  Section \ref{sec:conclusion} is our conclusion.
  
\section{The model}

The DGV model is a specific extra-dimensional theory featuring the brane as a topological defect: a kink-like domain-wall configuration \cite{rubshapdwbranes}.  Domain walls are stable classical solutions of suitable scalar field theories that exhibit a brane-like character, with energy-density peaked around the centre of the wall.  Unlike fundamental branes, they have a finite width, and are most naturally used to replace the $\delta$-function-like fundamental brane of the original RS2 model.  Like RS2, a 3+1d graviton zero mode is dynamically localized.  Unlike the original RS2 setup, all other degrees of freedom (fermions, scalars and gauge bosons) must be {\it dynamically} localized.  In contrast to the fundamental-brane case, it is not possible to simply postulate that various fields are confined to a domain-wall: one must have dynamics to do it, and that is the main challenge in developing realistic models of this kind.

The dynamical localization of chiral fermion zero-modes is automatic when 4+1d fermions Yukawa-couple to a background scalar field in the form of a kink \cite{rubshapdwbranes}.  Thus the chiral fermion structure of the SM can be naturally accommodated.  Similarly, additional scalar bosons such as a Higgs doublet can be dynamically localized to a domain wall through a Higgs potential that couples those extra scalars to the background scalar field configuration \cite{modetower}.  Such localized scalars can even obtain negative squared masses, thus triggering spontaneous symmetry breaking on the wall.

The most difficult issue is the localization of gauge bosons, with the need to maintain exact 3+1d gauge invariance to ensure gauge universality.  A promising mechanism was proposed by Dvali and Shifman \cite{dsmech}, the physics of which is quite different from the localization of fermions and scalars.  The idea is that a gauge group $G$ spontaneously breaks to a subgroup $H$ inside the wall, but is restored in the bulk.  The bulk gauge theory is taken to be in confinement phase.  The proposition is that the gauge bosons of $H$ are then dynamically localized to the wall as exactly massless states enjoying exact 3+1d gauge invariance.  There are two heuristic arguments for why this should be the case.  Dvali and Shifman themselves argued as follows: Take the case where $G = SU(2)$ and $H = U(1)$, and call the gauge boson of $U(1)$ the ``photon''.  The photon is obviously free to propagate as a massless gauge boson in the plane of the wall.  But in propagating transverse to the wall, into the bulk, the $SU(2)$ confinement regime is encountered, and the propagating states must be colorless and, importantly, massive glueballs.  The photon must incorporate itself into a massive glueball to enter the bulk.  But this mass gap makes this transition energetically disfavored, thus trapping the photon on the wall.  Subsequently, Arkani-Hamed and Schmaltz \cite{dualmeisnerrplusdsmech} presented another heuristic picture: the photon field lines must be repelled from the bulk, because a confinement-phase region is by definition unable to support diverging electric fields.  Thus the flux is channeled along the wall, effecting a dimensional reduction.  At large distances within the wall away from the source, the field lines exhibit 3+1d Coulomb form.  If the source is instead located in the bulk, then a flux tube leading to the wall is formed, with the field lines then diverging outward {\it as if} the source was located within the wall.  Thus, the long-distance behavior of the field lines within the wall is independent of where the source is placed.  If the source is smeared out along the extra dimension due to a profile function, then the corollary is that the asymptotic field line behavior is independent of the profile.  Charge universality is thus maintained, no matter how the source is distributed along the extra dimension.  These conclusions generalize to an arbitrary $G$ and $H$, provided any glueballs associated with $H$ (e.g.\ QCD glueballs) are less massive than $G$ glueballs. 

So, the way to develop potentially realistic domain-wall brane models is clear.  One first postulates a scalar field theory that admits a suitable topological domain wall solution.  This configuration must break gauge group $G$ to $H$, and $H$ must contain or be the SM gauge group.  Chiral fermion zero modes and additional scalars such as an electroweak Higgs doublet are then dynamically localized as sketched above.  The DGV model uses the minimal gauge structure where $G = SU(5)$ and $H = SU(3) \times SU(2) \times U(1)$ \cite{firstpaper}.

\subsection{The field content}

The scalar fields in the model are
\begin{gather}
\eta \sim 1, \\
\chi \sim 24, \\
\Phi \sim 5^*.
\label{eq:su5scalarreps}
\end{gather}
 The fermion content of the theory consists of the SM fermions with a gauge singlet right-handed neutrino for each generation $i=1,2,3$.
 The SM fermions are placed into the following $SU(5)$ representations,
\begin{gather}
\Psi^i_5 \sim 5^*, \\
\Psi^i_{10} \sim 10,
\label{eq:su5fermionreps}
\end{gather}
while the right handed neutrinos are singlets, 
\begin{equation}
N^i \sim 1.
\label{eq:rhneutrinosu5rep}
\end{equation}
$\Psi^i_5$ contains the charge conjugate of the right-chiral down-type quark and the left-chiral lepton doublet, and $\Psi^i_{10}$ contains the left-chiral quark doublet and the charge conjugates of the right-chiral up-type quark and electron-type lepton for the generation $i$. 

 In this $4+1d$ model, matter is confined to a domain-wall brane formed from a solitonic kink configuration for the $\eta$ field. 
 To implement the Dvali-Shifman mechanism, $SU(5)$ is broken inside the domain wall by the second background field $\chi$ which transforms under the adjoint representation. It attains a non-zero value for the hypercharge generator component inside the domain wall, so the gauge group respected on the domain wall is that of the SM. 

Chiral fermion zero modes are trapped on the domain wall through Yukawa interactions with the $\eta$ and $\chi$ background configurations. Similarly, additional scalar fields are trapped by introducing quartic interactions between those scalar fields and the background domain wall.

 For the purposes of this paper, we shall ignore gravity, although a similar analysis will have to be done with its inclusion in a later paper.  It has already been noted that the RS2 graviton localization mechanism also works for a domain-wall brane.  For a discussion of how gravity affects the dynamical localization of other fields, see Refs.~\cite{rsgravitydaviesgeorge2007, firstpaper}.

\subsection{The background domain wall configuration}

 The background domain wall configuration is formed from a self-consistent classical solution for the coupled fields $\eta$ and $\chi$. The singlet scalar field $\eta$ forms the kink-like domain wall, while the adjoint $\chi$ breaks $SU(5)$ down to $SU(3)_c\times{}SU(2)_L\times{}U(1)_Y$ on the domain wall by attaining a bump-like configuration.

 The relevant part of the action for describing the dynamics of the background is \cite{firstpaper},
\begin{equation}
S = \int{}d^5x (T-V_{\eta{}\chi{}}),
\label{eq:action}
\end{equation}
where T contains all the $SU(5)$ gauge-covariant kinetic terms for all the fields. $V_{\eta{}\chi{}}$ is the part of Higgs potential containing the quartic potentials for $\eta$ and $\chi$, with
\begin{equation}
V_{\eta{}\chi{}} = (c\eta{}^2-\mu^2_{\chi})Tr(\chi^2)+a\eta{}Tr(\chi^3)+\lambda_{1}[Tr(\chi^2)]^2+\lambda_{2}Tr(\chi^4)+l(\eta^2-v^2)^2.
\label{eq:backgroundpot1}
\end{equation}

 We want $\chi$ to break $SU(5)$ to the SM on the domain wall, while having the bulk respect the original gauge symmetry. We do this in the standard way by giving the component $\chi_1$ associated with the hypercharge generator $Y$ a non zero value on the brane, and having all the other components vanish. Thus the potential reduces to
\begin{equation}
V_{\eta{}\chi{}}=\frac{\tilde{\lambda}}{4}\chi^4_1+l(\eta^2-v^2)^2-\frac{1}{4}\sqrt{\frac{1}{15}}a\eta{}\chi_1^3+\frac{1}{2}(c\eta^2-\mu^2_{\chi})\chi^2_1,
\label{eq:backgroundpot2}
\end{equation}
where $\tilde{\lambda}=\lambda_1+\frac{7\lambda_2}{30}$. 

\begin{figure}[h]
\subfloat[]{\includegraphics[scale=0.7]{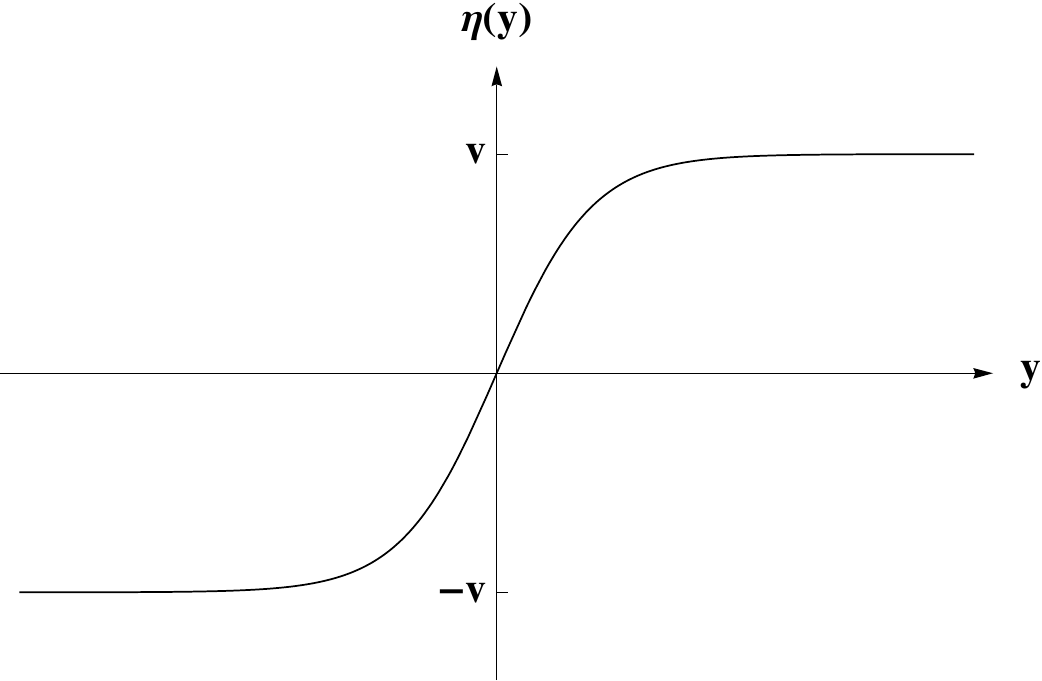}}
\subfloat[]{\includegraphics[scale=0.7]{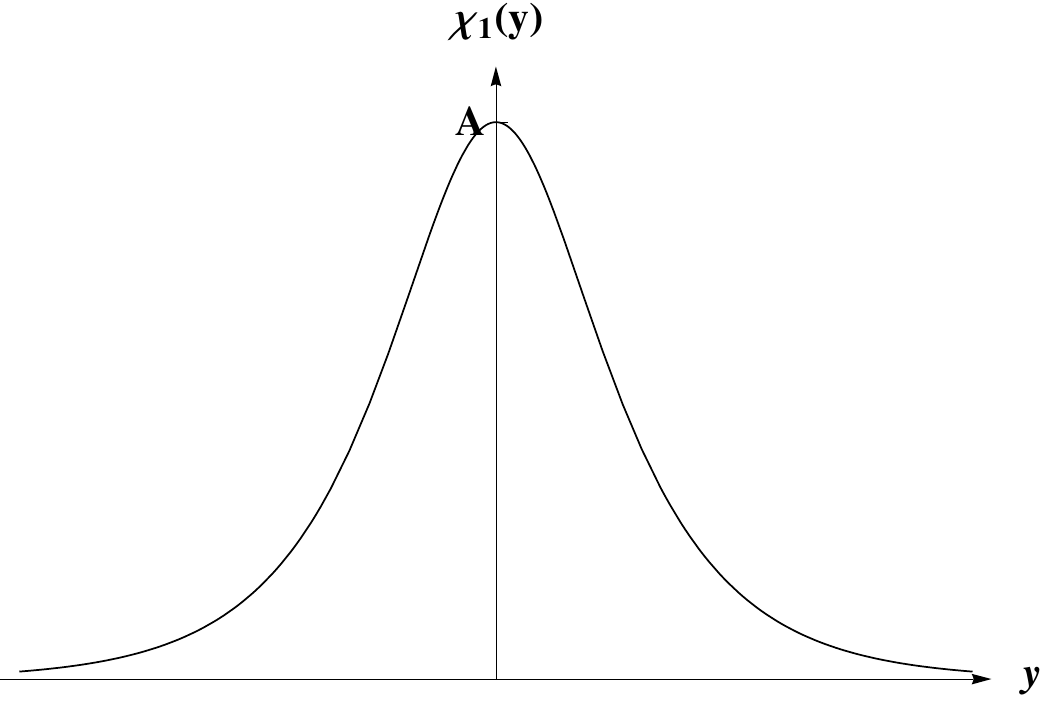}}
\caption{(a) The kink $\eta{}$ and (b) $\chi_1$ as functions of the extradimensional coordinate y}
\label{fig:background}
\end{figure}

 To find the background configuration, we need to solve the Euler-Lagrange equations for $\eta$ and $\chi_1$ subject to the boundary conditions
\begin{gather}
\eta{}(y=\pm{}\infty)=\pm{}v, \\
\chi_1{}(y=\pm{}\infty)=0,
\label{eq:backgroundboundaryconditions}
\end{gather}
which are degenerate global minima of $V_{\eta{}\chi}$. For the sake of simplicity, we choose to impose the constraints
\begin{gather}
2\mu^2_{\chi}(c-\tilde{\lambda})+(2c\tilde{\lambda}-4l\tilde{\lambda}-c^2)v^2=0, \\
a=0,
\label{eq:analyticsolconditions}
\end{gather}
yielding the analytic solutions,
\begin{equation}
\begin{gathered}
\eta{}(y) = v\tanh{(ky)}, \\
\chi_1(y) = A\sech{(ky)},
\end{gathered}
\label{eq:analyticsol}
\end{equation}
where $k^2=cv^2-\mu^2_{\chi}$, and $A^2=\frac{2\mu^2_{\chi}-cv^2}{\tilde{\lambda}}$. 
 We should stress that the above conditions are not fine tuning conditions, and they are chosen simply so that the background fields obtain analytic forms. To find solutions, these conditions need not be imposed, and for a finite range of parameters we can always find numerical solutions which are kink-like for $\eta$ and lump-like for $\chi$ \cite{firstpaper}.
 The graphs of these solutions for $\eta$ and $\chi_1$ are shown in Figures \ref{fig:background}(a) and \ref{fig:background}(b) respectively.

 The kink-like $\eta(y)$ has its energy density localized about $y=0$, forming the domain wall brane. The bump-like $\chi_1(y)$ breaks $SU(5)$ to $SU(3)_c\times{}SU(2)_L\times{}U(1)_Y$ on the domain wall. 

 To preserve the topological stability of the domain wall, a spontaneously broken $\mathbb{Z}_2$ reflection symmetry must be introduced. Under this discrete symmetry transformation,
\begin{equation}
\begin{gathered}
y \rightarrow{} -y, \\
\eta \rightarrow{} -\eta,\\
\chi \rightarrow{} -\chi, \\
\Psi^i_{5,10} \rightarrow{} i\Gamma^5\Psi^i_{5,10}, \\
N^i \rightarrow{} -i\Gamma^5N^i,
\end{gathered}
\label{eq:Z2symtransformations}
\end{equation}
where the 4+1d Gamma matrices $\Gamma^M$ for $M=0,1,2,3,5$ are defined 
\begin{equation}
\begin{gathered}
\Gamma^{\mu}=\gamma^{\mu}, \; \rm{for}\;M=\mu=0,1,2,3, \\
\Gamma^5 = -i\gamma^5.
\end{gathered}
\label{eq:5dgammamatrices}
\end{equation}
For the entirety of this paper, 4+1d Lorentz indices will be denoted with upper case Roman indices, while ordinary 3+1d Lorentz indices will be denoted with lower case Greek letters as usual. Also, $x^5=y$.
The next step is to localize the fermions to this background.

\subsection{Localizing the charged fermions and the left-chiral neutrino}

 To localize the charged fermions and the left-chiral neutrino to the domain wall, we need to couple them to the background. The relevant Yukawa Lagrangian which localizes the fermions, for one generation, is \cite{firstpaper}, 
\begin{equation}
Y_{DW} = h_{5\eta}\overline{\Psi}_5\Psi_5\eta+h_{5\chi}\overline{\Psi}_5\chi{}\Psi_5 +h_{10\eta}Tr(\overline{\Psi}_{10}\Psi_{10})\eta-2h_{10\chi}Tr(\overline{\Psi}_{10}\chi{}\Psi_{10}).
\label{eq:Ydw}
\end{equation}

 The resulting 5d Dirac equation, for the charged fermions, is
\begin{equation}
i\Gamma^M\partial_M\Psi_{nY}(x,y)-h_{n\eta}\eta(y)\Psi_{nY}(x,y)-\sqrt{\frac{3}{5}}\frac{Y}{2}\chi_1(y)\Psi_{nY}(x,y)=0, \;n=5,10.
\label{eq:5dlocdiraceq}
\end{equation}

As explained in \cite{firstpaper}, to find the zero modes, it is enough to look for solutions for each charged fermion of the form $\Psi_{nY}(x,y)=f_{nY}(y)\psi_{nY}(x)$ where the $\psi_{nY}(x)$ are 3+1d massless, left-chiral spinor fields. Substituting this into the above Dirac equations yields the solutions for the profiles,
\begin{equation}
f_{nY}(y) = C_{nY}e^{-b_{nY}(y)}, \; \rm{for} \; n=1,5,10,
\label{eq:profilesol}
\end{equation}
where the $C_{nY}$ are normalisation constants, and
\begin{equation}
\begin{gathered}
b_{nY}(y) = \tilde{h}_{n\eta}\log{(\cosh{(ky)}})+Y\sqrt{\frac{3}{5}}\tilde{h}_{n\chi}\arctan{\left(\tanh{(\frac{ky}{2})}\right)}, \\
\tilde{h}_{n\eta} = \frac{h_{n\eta}v}{k}, \\
\tilde{h}_{n\chi} = \frac{h_{n\chi}A}{k}.
\end{gathered}
\label{eq:logprof}
\end{equation}
 These profiles have maxima at 
\begin{equation}
y_{max,nY}=\frac{1}{k}\arcsinh{\left(-\sqrt{\frac{3}{5}}\frac{Y}{2}\frac{\tilde{h}_{n\chi}}{\tilde{h}_{n\eta}}\right)}.
\label{eq:profilemax}
\end{equation}
 Hence, the charged fermions and the left-chiral neutrino, which reside in the non-trivial representations of $SU(5)$, get split along the extra-dimension according to their hypercharges and Yukawa couplings to the background. A similar effect was used in \cite{kakizakidoublettriplet, kakizakimassfitting}.

 During the mass fitting sections, we will need to describe the theory in terms of non-dimensionalized variables and profiles, since we do not know the value of $k$. The non-dimensionalized domain-wall Yukawa couplings $\tilde{h}_{n\eta}$ $\tilde{h}_{n\chi}$ have already been defined in Eq. \ref{eq:logprof} and so we just need to non-dimensionalise the profiles. Defining the non-dimensionalized extra-dimensional coordinate, $\tilde{y}$, as
\begin{equation}
\tilde{y}=ky,
\label{eq:dimensionlesscoordinate}
\end{equation}
and changing variables, we see that the normalisation condition for the profiles becomes 
\begin{equation}
\int{}f_{nY}(\tilde{y})^{\dagger}f_{ny}(\tilde{y})\;d\tilde{y} = k.
\label{eq:normalisationinytilde}
\end{equation}
Hence, in order to use functions which are normalised to one over $\tilde{y}$, we define the non-dimensionalized profiles, $\tilde{f}_{nY}(\tilde{y})$, as
\begin{equation}
\tilde{f}_{nY}(\tilde{y}) = k^{-\frac{1}{2}}f_{nY}(\tilde{y}).
\label{eq:dimensionlessprofiles}
\end{equation}

 Thus, the profiles $f_{nY}$ scale as $k^{\frac{1}{2}}$ times a dimensionless function, which is not surprising since we know that the 4+1d field $\Psi_{nY}(x,y)$ has mass dimension $2$, while the 3+1d field $\psi_{nY}(x)$ has mass dimension $\frac{3}{2}$ as usual.

 Since the factor of $k$ in Eq. \ref{eq:normalisationinytilde} will always arise in the normalisation condition when changing variables from $y$ to $\tilde{y}$, we will define the non-dimensionalised profiles for any field in the same way as Eq. \ref{eq:dimensionlessprofiles}, and they will be denoted with the same symbol used for the dimensionful profiles but with an overscript tilde. 

 In the case that we have $m>1$ generations of fermions, $Y_{DW}$ is generalized to
\begin{equation}
Y_{DW} = h^{ij}_{5\eta}\overline{\Psi^i_5}\Psi^j_5\eta+h^{ij}_{5\chi}\overline{\Psi^i_5}\chi{}\Psi^j_5 +h^{ij}_{10\eta}Tr(\overline{\Psi^i_{10}}\Psi^j_{10})\eta-2h^{ij}_{10\chi}Tr(\overline{\Psi^i_{10}}\chi{}\Psi^j_{10}),
\label{eq:Ydw3gen}
\end{equation}
where $i$ and $j$ are summed from $1$ to $m$. Hence, in the general case, there can be intergenerational mixing between the quarks and leptons through the interaction with the background. The background couplings $h_{n\eta}$ and $h_{n\chi}$ have now become $3\times{}3$ Hermitian matrices over flavour space and need not commute. To solve the equations, we look for zero mode solutions of the form,
\begin{equation}
\Psi_{nY}^i(x,y)=f_{nY}^{ij}(y)\psi_{nY}^j(x),
\label{3genprofileansatz}
\end{equation}
where the $\psi_{nY}^j(x)$ are massless left chiral 3+1d fields for $n=5,10$. Putting this into the 4+1d Dirac equation results in the matrix differential equation for the profiles $f_{nY}$, which are now $3\times{}3$ matrix valued functions of $y$,
\begin{equation}
\frac{df_{nY}(y)}{dy}+\eta(y)h_{n\eta}f_{nY}(y)+\sqrt{\frac{3}{5}}\frac{Y}{2}\chi_1(y)h_{n\chi}f_{nY}(y)=0.
\label{eq:profilematrixeq}
\end{equation}
 The case where $h_{n\eta}$ and $h_{n\chi}$ do not commute, which leads to a natural realisation of the $twisted$ split fermion scenario discussed in Refs. \cite{twistedsplitfermions, cpandtwistedsf}, cannot be solved analytically, and so for the sake of simplicity we will only search the parameter space that obeys,  
\begin{equation}
[h_{n\eta},h_{n\chi}]=0, \; \rm{for}\;n=5,10.
\label{eq:vanishingcommutatorchoice}
\end{equation}
Since both the matrices are required to be Hermitian as well, they are thus simultaneously diagonalizable, so that for some unitary matrices $S_n$, 
\begin{equation}
\begin{gathered}
S_nh_{n\eta}S^{\dagger}_n = diag(h^1_{n\eta}, h^2_{n\eta}, \cdots{}, h^m_{n\eta}), \\
S_nh_{n\chi}S^{\dagger}_n = diag(h^1_{n\chi}, h^2_{n\chi}, \cdots{}, h^m_{n\chi}),
\end{gathered}
\label{eq:domainwallmatrixdiagbasis}
\end{equation}
where the $h^i_{n\eta}$ and $h^i_{n\chi}$ are understood to be the eigenvalues of $h_{n\eta}$ and $h_{n\chi}$ respectively. Choosing to localize left-chiral zero modes for $\Psi^i_{5,10}$ is then equivalent to demanding that all the eigenvalues of $h_{n\eta}$ are positive definite. Solving the 5d Dirac equation then yields the general solution for the profiles, 
\begin{equation}
\begin{gathered}
f_{nY}(y) = S^{\dagger}diag(C^1_{nY}e^{-b^{1}_{nY}(y)}, C^2_{nY}e^{-b^{2}_{nY}(y)}, \cdots{}, C^m_{nY}e^{-b^{m}_{nY}(y)})V_{nY}, \\
b^{i}_{nY}(y) = \tilde{h}^i_{n\eta}\log{(\cosh{(ky)}})+Y\sqrt{\frac{3}{5}}\tilde{h}^i_{n\chi}\arctan{\left(\tanh{(\frac{ky}{2})}\right)}
\end{gathered}
\label{eq:3genprofileforvancomm}
\end{equation}
 Here we have written the multi-generation solutions in terms of the solutions for the one generation case. The $C^i_{nY}$ are normalisation constants, chosen such that the profile matrix $f_{nY}(y)$ satisfies the normalisation condition, 
\begin{equation}
\int{}f^{\dagger}_{nY}(y)f_{nY}(y)\;dy = \mathbbm{1}.
\label{eq:3genprofnormalisation}
\end{equation}
 The parameters $\tilde{h}^i_{n\eta}$ and $\tilde{h}^i_{n\chi}$ are the non-dimensionalized versions of $h^i_{n\eta}$ and $h^i_{n\chi}$, and are defined in the same way as the non-dimensionalized constants from the one generation case were in Eq. \ref{eq:logprof}. The $V_{nY}$ are unitary matrices which are present since the solution is unique up to matrix multiplication. The $V_{nY}$, in fact, correspond to a choice of which 3+1d states are the domain wall eigenstates and thus localized to the wall. Unless otherwise stated, we will assume these to be the same as the weak interaction eigenstates.

\subsection{Adding singlet right-handed neutrinos}

 To localize the right-chiral neutrinos, we need to couple them to the background. As they are gauge singlets, they cannot couple to the adjoint Higgs $\chi$. Thus we can only add,
\begin{equation}
-h^{ij}_{1\eta}\overline{N^i}N^j\eta{},
\label{eq:rhneutrinoYdw}
\end{equation}
to $Y_{DW}$. The relative minus sign in front of the Yukawa interactions for the $N^i$ is introduced because for these fields we want localized right-chiral zero modes which represent the right-handed neutrinos in the effective 3+1d theory, as opposed to left-chiral zero modes. This allows us to treat $h_{1\eta}$ in the same way as $h_{5\eta}$ and $h_{10\eta}$. 

 Writing down the 5d Dirac equation for the $N^i$, and demanding that $N^i(x, y)=f_N^{ij}(y)N^j(x)$, where the $N^j(x)$ are 3+1d right chiral zero modes, in similar fashion to the charged fermions, leads to the profile,
\begin{equation}
f_N(y) = S^{\dagger}_1diag(C^1_{1}e^{-\frac{h^{1}_{1\eta}v}{k}\log{(\cosh{(ky)})}}, C^2_1e^{-\frac{h^{2}_{1\eta}v}{k}\log{(\cosh{(ky)})}}, \cdots, C^m_1e^{-\frac{h^{m}_{1\eta}v}{k}\log{(\cosh{(ky)})}})V_N,
\label{eq:rhneutrinoprofiles}
\end{equation}
where $S_1$ is again a choice of basis matrix for the 5d fields, $V_N$ is a change of basis matrix for the 4d fields, the $C^i_1$ are normalisation constants, and the $h^i_{1\eta}$ are the positive definite eigenvalues of the Yukawa matrix $h_{1\eta}$.

\subsection{Localizing the electroweak symmetry breaking Higgs boson}
\label{sec:higgsloc}

 The electroweak breaking Higgs doublet is localized in a very similar manner to the fermions. The most general localizing Higgs potential which respects the $SU(5)$ and discrete symmetries is 
\begin{equation}
V_{\Phi} = \mu^2_{\Phi}\Phi^{\dagger}\Phi+\lambda_3(\Phi^{\dagger}\Phi)^2+\lambda_4\Phi^{\dagger}\Phi{}\eta^2+2\lambda_5\Phi^{\dagger}\Phi{}Tr[\chi^2]+\lambda_6\Phi^{\dagger}(\chi^T)^2\Phi+\lambda_7\Phi^{\dagger}\chi^T\Phi{}\eta.
\label{eq:higgspotential}
\end{equation}
To find the profiles of the electroweak Higgs doublet, $\Phi_w$, and the colored Higgs triplet, $\Phi_c$, embedded in the quintet $\Phi$, we search for solutions of the form,
\begin{equation}
\Phi_{w,c}(x,y) = p_{w,c}(y)\phi_{w,c}(x),
\label{eq:higgsprofileansatz}
\end{equation}
where the $p_{w,c}$ are the respective profiles, and $\phi_{w,c}$ satisfy the Klein-Gordon equations,
\begin{equation}
\boxempty{}_{3+1d}\phi_{w,c}=m^2_{w,c}\phi_{w,c}+\dots{}
\label{eq:higgskkmode4dKGeq}
\end{equation}
where $m_{w,c}$ are the masses of the lowest energy modes for $\Phi_{w,c}$. Substituting this ansatz into the 4+1d KG equation with the potential $V_{\Phi}$, one obtains the equations for the profiles
\begin{equation}
-\frac{d^2p_{w,c}}{dy^2}+W_Y(y)p_{w,c}(y)=m^2_{w,c}p_{w,c}(y),
\label{eq:higgsprofileSE}
\end{equation}
where 
\begin{equation}
W_Y(y)=\mu^2_{\Phi}+\lambda_4\eta^2+\lambda_5\chi^2_1+\frac{3Y^2}{20}\lambda_6\chi^2_1+\sqrt{\frac{3}{5}}\frac{Y}{2}\lambda_7\eta{}\chi_1.
\label{eq:effectivehiggslocpot}
\end{equation}

 Changing variables to the dimensionless coordinate $\tilde{y}$ defined in Eq. \ref{eq:dimensionlesscoordinate}, the potentials of the above Schr\"{o}dinger equations can be rewritten as shifted hyperbolic Scarf potentials, that is we can write them in the form
\begin{eqnarray}
 & \left[-\frac{d^2}{d\tilde{y}^2}+A^2_Y+(B^2_Y-A^2_Y-A_Y)\sech{(\tilde{y})}^2 + B_Y(2A_Y+1)\sech{(\tilde{y})} \tanh{(\tilde{y})}\right]p_{w,c}(\tilde{y}) & \nonumber\\
& =  \lambda_{w,c}p_{w,c}(\tilde{y}) &
\label{eq:hyperbolicscarfpotform}
\end{eqnarray}
where
\begin{equation}
\begin{gathered}
A_Y = \frac{-1+\sqrt{2((\tilde{\lambda}_5+\frac{3Y^2}{20}\tilde{\lambda}_6-\tilde{\lambda}_4-\frac{1}{4})^2+\frac{3Y^2}{20}\tilde{\lambda}^2_7)^{\frac{1}{2}}-2\tilde{\lambda}_5-\frac{3Y^2}{10}\tilde{\lambda}_6+2\tilde{\lambda}_4+\frac{1}{2}}}{2}, \\
B_Y = \frac{\sqrt{\frac{3}{5}}\frac{Y}{2}\tilde{\lambda}_7}{\sqrt{2((\tilde{\lambda}_5+\frac{3Y^2}{20}\tilde{\lambda}_6-\tilde{\lambda}_4-\frac{1}{4})^2+\frac{3Y^2}{20}\tilde{\lambda}_7^2)^{\frac{1}{2}}-2\tilde{\lambda}_5-\frac{3Y^2}{10}\tilde{\lambda}_6+2\tilde{\lambda}_4+\frac{1}{2}}},
\end{gathered}
\label{eq:hypscarfeffectivebackgroundcouplings}
\end{equation}
and the non-dimensionalized Higgs parameters and masses are defined as
\begin{equation}
\begin{gathered}
\tilde{\lambda}_4 = \frac{\lambda_4v^2}{k^2}, \\
\tilde{\lambda}_5 = \frac{\lambda_5A^2}{k^2}, \\
\tilde{\lambda}_6 = \frac{\lambda_6A^2}{k^2}, \\
\tilde{\lambda}_7 = \frac{\lambda_7vA}{k^2}, \\
\tilde{\mu}^2_\Phi = \frac{\mu^2_\Phi}{k^2}, \\
\tilde{m}^2_{w,c} = \frac{m^2_{w,c}}{k^2},
\end{gathered}
\label{eq:dimensionlesshiggsparameters}
\end{equation}
and $\lambda_{w,c}=\tilde{m}^2_{w,c}-\tilde{\mu}^2_{\phi}-\tilde{\lambda}_4+A^2_Y$ are the eigenvalues of the equations for the electroweak Higgs and the colored Higgs hyperbolic Scarf potentials. 

 The hyperbolic Scarf potential has been well studied \cite{castillohyperscarfpot} as it is a member of a class of potentials satisfying the shape-invariance condition in supersymmetric quantum mechanics (for more on shape-invariant potentials see \cite{shapeinvkharesukdab, levaishapeinvariantpots}). For $A_Y>0$, it is known to have a set of discrete bound modes for $n=0, 1, ..., \lfloor{}A_Y\rfloor{}$, with eigenvalues
\begin{equation}
\lambda^n_{w,c} = 2nA_Y-n^2.
\label{eq:hyperbolicscarfeigenvalues}
\end{equation}
Combining this with the previous equations for $\lambda_{w,c}$, we see that the potentials have a discrete set of bound modes with masses given by
\begin{equation}
\tilde{m}^2_{n,w,c} = \tilde{\mu}^2_{\Phi}+\tilde{\lambda}_4-(A_Y-n)^2.
\label{eq:higgsKKmodemasses}
\end{equation}
 The physical electroweak Higgs and colored Higgs fields in the effective 4d theory on the brane correspond to the $n=0$ modes, and they thus exist in the 4d theory if $A_Y>0$. Assuming this, the profiles for these Higgs particles, $p_w(y)$ and $p_c(y)$ respectively, have the same form as those of the zero mode profiles for the charged fermions, 
\begin{equation}
\begin{gathered}
p_{w,c}(y) = C_{w,c}e^{-b_{w,c}(y)}, \\
b_{w,c}(y) = A_Y\log{(\cosh{(ky)}})+2B_Y\arctan{\left(\tanh{(\frac{ky}{2})}\right)}.
\end{gathered}
\label{eq:higgsprofilesolutions}
\end{equation}
Hence, we can interpret $A_Y$ and $B_Y$ to be effective couplings of the Higgs fields to the kink and the lump respectively.

 The effective couplings $A_Y$ and $B_Y$ depend on the hypercharges, and thus they are in general different for the two Higgs components. This has a number of consequences. Firstly, since the masses of the electroweak and colored Higgs depend on their respective $A_Y$, the masses of the two components are split. There exists a parameter region where the electroweak Higgs has a tachyonic mass, $m^2_w<0$, while that for the colored Higgs (if a bound state exists) is non-tachyonic, thus inducing electroweak symmetry breaking on the brane while preserving $SU(3)_c$, as is desired. Since we know the exact form of the masses, a straightforward analysis shows that this parameter region is
\begin{equation}
A^2_{+2/3}<\tilde{\mu}^2_\Phi+\tilde{\lambda}_4<A^2_{-1}.
\label{eq:electroweaksymbreakingregion}
\end{equation}

 Secondly, as there only exist discrete bound modes for a species if $A_Y>0$, there exist parameter regions where the electroweak Higgs component will have discrete bound modes localized to the domain wall while at the same time the colored Higgs will have only unbound continuum modes in its spectrum. This suggests that an alternate approach to suppressing colored Higgs induced proton decay may be possible, as the continuum modes propagate in the full 4+1d spacetime so that the partial width contributed to proton decay from these modes may be suppressed by further powers of $M_{GUT}$. The analysis of this situation is beyond the scope of this paper.

 Note that it is possible for more than one KK excitation of the Higgs doublet to have nonzero vacuum expectation values, thus naturally generating a multi-Higgs doublet model on the brane. However, for simplicity, we will choose parameters such that only the electroweak Higgs has a tachyonic mass, and not its KK modes, and we will also have a bound state for the colored Higgs. For the purposes of this paper, we will use three such choices. 

\begin{figure}[h]
\begin{center}
\includegraphics[scale=0.7]{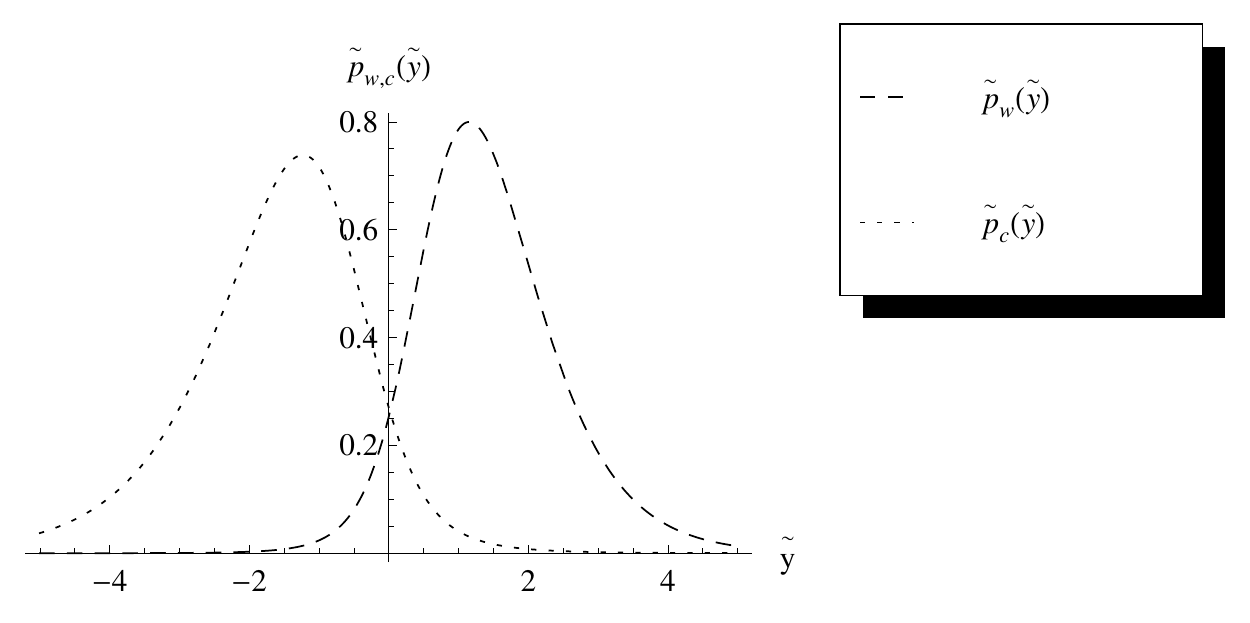}
\caption{Profiles for the colored Higgs, $p_c$, and electroweak Higgs, $p_w$, for the parameter choices given in Eq.\ref{eq:firsthiggschoice}.}
\label{fig:firsthiggsprofilegraphs}
\end{center}
\end{figure}

 For the first choice, 
\begin{equation}
\begin{gathered}
\mu^2_{\Phi}=k^2, \\ 
\lambda_4=\frac{0.5k^2}{v^2}, \\ 
\lambda_5=\frac{k^2}{A^2},  \\
\lambda_6=\frac{k^2}{A^2},  \\
\lambda_7=\frac{20k^2}{vA},
\end{gathered}
\label{eq:firsthiggschoice}
\end{equation}
 the mass eigenvalues are $m^2_w=-0.510k^2$, and $m^2_c=0.380k^2$. The graphs of the profiles are shown in Fig. \ref{fig:firsthiggsprofilegraphs}.

\begin{figure}[h]
\begin{center}
\includegraphics[scale=0.7]{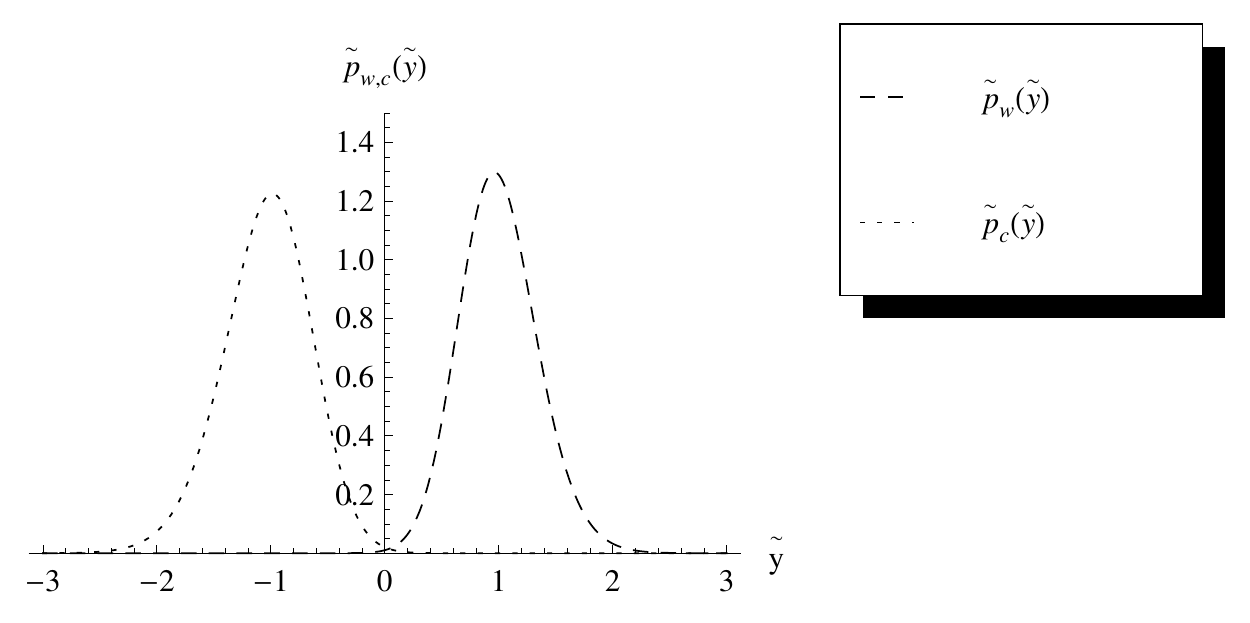}
\caption{Higgs profiles for parameter choices given in Eq.\ref{eq:secondhiggschoice}.}
\label{fig:secondhiggsprofilegraphs}
\end{center}
\end{figure}

 For the second choice, 
\begin{equation}
\begin{gathered}
\mu^2_{\Phi}=65k^2, \\
\lambda_4=\frac{0.5k^2}{v^2}, \\ 
\lambda_5=\frac{10k^2}{A^2}, \\ 
\lambda_6=\frac{10k^2}{A^2} \\ 
\lambda_7=\frac{500k^2}{vA},
\end{gathered}
\label{eq:secondhiggschoice}
\end{equation}
 the mass eigenvalues are $m^2_w=-16.8k^2$, and $m^2_c=13.2k^2$. As can be seen in Fig. \ref{fig:secondhiggsprofilegraphs}, this leads to profiles which are much more localized than those for the first choice of parameters. As we will see, this has important consequences for the spread of domain wall Yukawa couplings and for the suppression of some of the decay modes for colored Higgs induced proton decay.

 For the third choice, 
\begin{equation}
\begin{gathered}
\mu^2_{\Phi}=97700k^2, \\
\lambda_4=\frac{-75000k^2}{v^2}, \\
 \lambda_5=\frac{15000k^2}{A^2}, \\
\lambda_6=\frac{-750000k^2}{A^2} \\
\lambda_7=\frac{20000k^2}{vA}. 
\end{gathered}
\label{eq:thirdhiggschoice}
\end{equation}
 The resultant squared-masses for the lowest energy modes are $m^2_w=-296k^2$ and $m^2_c=2.25\times{}10^{4}k^2$.
\begin{figure}[h]
\begin{center}
\includegraphics[scale=0.7]{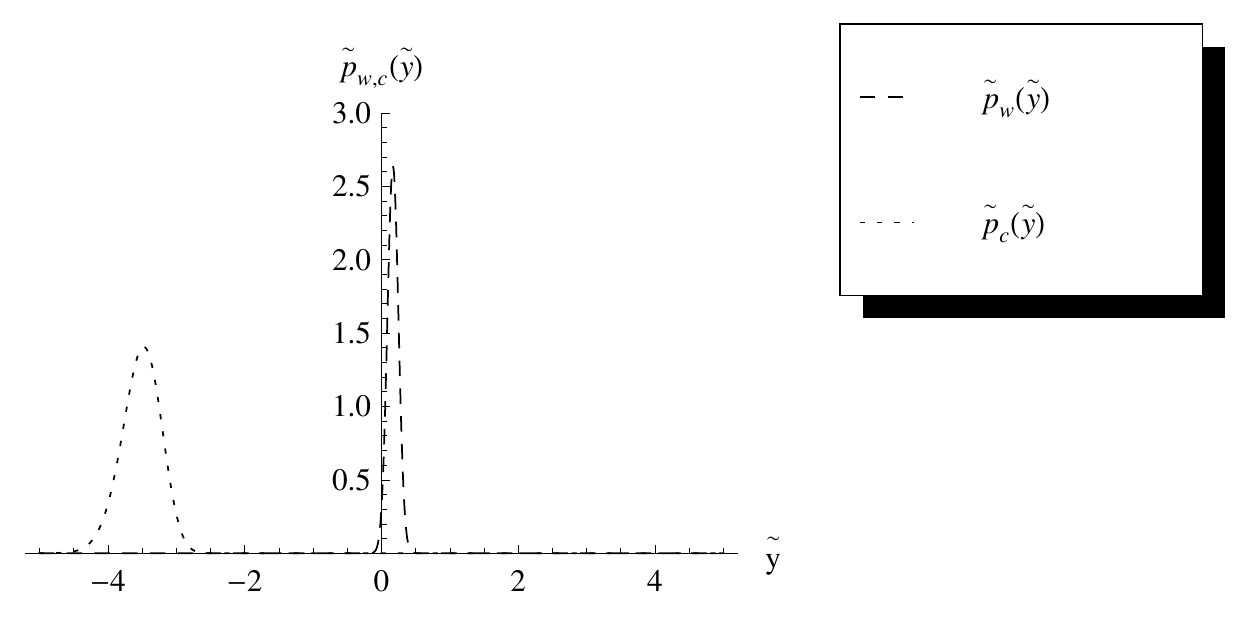}
\caption{Higgs profiles for parameter choices given in Eq.\ref{eq:thirdhiggschoice}.}
\label{fig:thirdhiggsprofilegraphs}
\end{center}
\end{figure}
As we can see in the graphs of the profiles in Fig. \ref{fig:thirdhiggsprofilegraphs}, for this parameter region, the electroweak Higgs is highly peaked near the brane at $y=0$, while the colored Higgs is more delocalized and substantially displaced from the wall. This parameter choice exploits the property of the Higgs sector that effective kink and lump couplings $A_Y$ and $B_Y$ are not the same for the colored and electroweak Higgs. As we will see, this kind of parameter choice can lead to suppression of all decay modes for colored Higgs induced proton decay, and ensure that the partial lifetimes for these modes are all many orders of magnitude above the current lower bounds.

 Note that the Higgs vacuum expectation value, $\langle{}\phi_w\rangle{}$, is not uniquely determined by the constants which determine the Higgs profile. By dimensional reduction of the action,
\begin{equation}
S = \int{}d^5x\;(\partial^M\Phi{})^{\dagger}(\partial_M\Phi)-V_\Phi,
\label{eq:5dhiggsaction}
\end{equation}
one can show that the effective electroweak symmetry breaking potential is 
\begin{equation}
V_{EW}(\phi_w) = \lambda'(\phi_w^{\dagger}\phi_w)^2+m^2_w\phi^{\dagger}_w\phi_w,
\label{eq:4dewsymbreakpotential}
\end{equation}
where
\begin{equation}
\lambda' = \lambda_3\int{}p^4_{w}(y)\;dy.
\label{eq:effectivequarticcoupling}
\end{equation}
Thus the VEV of the Higgs doublet is 
\begin{equation}
\begin{gathered}
\langle{}\phi_w\rangle{} = \sqrt{\frac{-m^2_w}{2\lambda'}}  \\
                         = \sqrt{\frac{-m^2_w}{2\lambda_3\int{}p^4_{w}(y)\;dy}},
\end{gathered}
\label{eq:higgsvev}
\end{equation}
and so whatever we choose for the other constants, we can always adjust $\lambda_3$ appropriately so that we get the correct VEV of 174 GeV.

\subsection{Generating mass matrices for the charged fermions}

 The electroweak Yukawa Lagrangian, $Y_5$,  from \cite{firstpaper} which generates masses for the charged fermions is generalized to 
\begin{equation}
Y_5 = h^{ij}_{-}\overline{(\Psi^i_5)^C}\Psi^j_{10}\Phi+h^{ij}_{+}\epsilon^{\alpha{}\beta{}\gamma{}\delta{}\kappa{}}\overline{(\Psi^i_{10})^C}_{\alpha{}\beta{}}\Psi^j_{10\gamma{}\delta{}}\Phi^{*}_{\kappa}+h.c
\label{eq:ewyukawalagrangian},
\end{equation}
for $m$ generations of fermions. Here, lower case Greek letters are $SU(5)$ indices and the lower case Roman letters indicate flavor. 

 The $h_{-}$ terms generate mass matrices for the down-type quarks and electron type leptons, while the $h_{+}$ terms generate a mass matrix for the up-type quarks. Extracting the components from each term which generate 3+1d masses and performing dimensional reduction, one finds the mass matrices to be
\begin{equation}
\begin{gathered}
M_u = 4\overline{v}\int{}f^{\dagger}_{u_R}(y)h_{+}f_{Q}(y)p_w(y)\;dy, \\
M_d = \frac{1}{\sqrt{2}}\overline{v}\int{}f^{\dagger}_{d_R}(y)h_{-}f_{Q}(y)p_w(y)\;dy, \\
M_e = \frac{1}{\sqrt{2}}\overline{v}\int{}f^{\dagger}_{e_R}(y)h_{-}f_{l}(y)p_w(y)\;dy,
\end{gathered}
\label{eq:chargedfermionmassmatrices}
\end{equation}
where $p_w(y)$ is the profile of the electroweak Higgs doublet which is embedded in $\Phi$, and $\overline{v}=174\;GeV$ is the vacuum expectation value of the electroweak Higgs field attained on the brane. 

 Converting to dimensionless quantities, and defining the non-dimensionalized electroweak Yukawa couplings by
\begin{equation}
\begin{gathered}
\tilde{h}_{+} = k^{\frac{1}{2}}h_{+}, \\
\tilde{h}_{-} = k^{\frac{1}{2}}h_{-},
\end{gathered}
\label{eq:dimensionlesschargedsecewyukawas}
\end{equation}
we see that these mass matrices can be rewritten as 
\begin{equation}
\begin{gathered}
M_u = 4\overline{v}\int{}\tilde{f}^{\dagger}_{u_R}(y)\tilde{h}_{+}\tilde{f}_{Q}(y)\tilde{p}_w(y)\;d\tilde{y}, \\
M_d = \frac{1}{\sqrt{2}}\overline{v}\int{}\tilde{f}^{\dagger}_{d_R}(y)\tilde{h}_{-}\tilde{f}_{Q}(y)\tilde{p}_w(y)\;d\tilde{y}, \\
M_e = \frac{1}{\sqrt{2}}\overline{v}\int{}\tilde{f}^{\dagger}_{e_R}(y)\tilde{h}_{-}\tilde{f}_{l}(y)\tilde{p}_w(y)\;d\tilde{y},
\end{gathered}
\label{eq:chargedfermionmassmatricesindimensionlessvars}
\end{equation}

 There are some important consequences of the above forms of the mass matrices, which depend on overlap integrals of the profiles for the left and right chiral fermions and the electroweak Higgs. Firstly, the overlap integral dependence means we avoid the usual incorrect mass relations like $m_e=m_d$ which are characteristic of ordinary 3+1d $SU(5)$ models with a Higgs quintet. This is also the reason why we do not need a Higgs belonging to the $45$ representation of $SU(5)$ containing an electroweak Higgs triplet to get the Georgi-Jarlskog relations \cite{georgijarlskog}. 
 Thirdly, since the fermions are split according to their hypercharges, and the splittings are dependent on the background couplings, we can potentially generate the fermion mass hierarchy and mixings by splitting the fermions appropriately so that the overlap integrals are in the desired ratios. It will be shown in a later section that this can be done.

\subsection{Generating Dirac neutrino masses}

 To generate Dirac masses for the neutrinos, we need to add Yukawa interactions involving the $\Psi^i_5$, which contain the left handed neutrinos, the $N^i$, which contain the right handed neutrinos, and $\Phi$ which contains the electroweak Higgs. The correct terms to add to $Y_5$ which are both $SU(5)$ invariant and respect the reflection symmetry which preserves the topological stability of the domain wall are 
\begin{equation}
(h^{\dagger}_3)^{ij}\overline{\Psi^i_5}\Phi{}N^j+h.c.
\label{eq:rhneutrinoewyukawalagrangian}
\end{equation}

 Reducing these terms to their SM components, and integrating out the extra-dimensional dependence, one finds the resulting Dirac mass matrix for the neutrinos to be
\begin{equation}
m_{\nu}=\overline{v}\int{}f^{\dagger}_{\nu_R}(y)h_3f_{l}(y)p_w(y)\;dy.
\label{eq:diracneutrinomassmatrix}
\end{equation}
Defining the dimensionless neutrino Yukawa couplings as 
\begin{equation}
\tilde{h}_3 = k^{\frac{1}{2}}h_3,
\end{equation}
and changing to non-dimensionalized quantities, we can rewrite the Dirac mass matrix for the neutrino as
\begin{equation}
m_{\nu}=\overline{v}\int{}\tilde{f}^{\dagger}_{\nu_R}(y)\tilde{h}_3\tilde{f}_{l}(y)\tilde{p}_w(y)\;d\tilde{y}.
\label{eq:diracneutrinomassmatrixinnondimvars}
\end{equation}

\subsection{Generating Majorana neutrino masses}

 Let us consider one generation first. To generate a Majorana mass for the neutrino, we need to add terms to the Lagrangian that will dimensionally reduce to terms proportional to $\overline{\nu^c_R}\nu_R$ in the effective 4d theory. Thus, we might want to consider adding a term like
\begin{equation}
\overline{N}N^C+h.c.
\label{eq:majoranamassgeneratingterm}
\end{equation}
This is obviously gauge invariant, and it turns out that it is also invariant under the discrete reflection symmetry as well. We first need to consider what implications the addition of this term has for the existence of solutions of the 5d Dirac equation. The relevant Lagrangian is 
\begin{equation}
L_{N,DW} = i\overline{N}\Gamma^M\partial_MN+h_{1\eta}\overline{N}N\eta-\frac{1}{2}m(\overline{N}N^C+\overline{N^C}N),
\label{eq:5dloclagrangianwithmajmass}
\end{equation}
and thus the 5d Dirac equation becomes
\begin{equation}
i\Gamma^M\partial_MN+h_{1\eta}N\eta-mN^C=0.
\label{eq:5dlocdiraceqwithmajmass}
\end{equation}
Demanding the conditions that 
\begin{equation}
\begin{gathered}
N(x, y) = f_N(y)\nu_R(x), \\
\gamma^5\nu_R = \nu_R, \\
i\gamma^\mu\partial_\mu\nu_R = m'(\nu_R)^c,
\end{gathered}
\label{eq:majrhneutrinoprofileansatz}
\end{equation}
and noting that the parts proportional to $\nu_R$ and $(\nu_R)^c$ must be independent of each other as the corresponding spinors transform as right-chiral and left-chiral spinors respectively, we get two independent equations for $f_N$,
\begin{equation}
\begin{gathered}
\frac{df_N}{dy}+h_{1\eta}\eta(y)f_N(y) = 0, \\
m'f_N(y)-imf^*_N(y) =0.
\end{gathered}
\label{eq:majrhneutrinoprofileeq}
\end{equation}
The first of the equations above is exactly the same differential equation as before without the new term, and thus the $f_N$ must also have the same form as before, 
\begin{equation}
f_N(y) = C_Ne^{-\frac{h_{1\eta}v}{k}\log{(\cosh{(ky)})}}.
\label{eq:majrhneutrinoprofile}
\end{equation}
The second condition then implies that $m'=|m|$, and since any phase can just be absorbed into the definition of $N$, we can take $m'=m$. Hence, instead of a right-chiral zero mode, we now have a right-chiral Majorana mode of mass $m$ localized to the domain wall.

 Similarly with three generations, the profiles are unaltered by the Majorana mass terms, and the 3+1d Majorana mass matrix after dimensional reduction is then
\begin{equation}
M_{Maj, 3+1d} = \int{}f^T(y)m_{4+1d}f(y)\;dy.
\label{eq:neutrinomajoranamassmatrix}
\end{equation}

 We have thus successfully shown that both Dirac and Majorana masses can be generated with the addition of a right chiral singlet neutrino, and thus the see-saw mechanism can be employed. We will now demonstrate that the fermion mass hierarchy and small CKM mixing angles can be generated from split fermion idea.

\section{Generating the flavor hierarchy and mixing angles}
\label{sec:hierarchy}

 The fermion mass matrices depend on overlap integrals of the fermion profiles and the electroweak Higgs. Since the left-chiral and right-chiral components are naturally split according to their hypercharges, and since these overlap integrals are exponentially sensitive to these splittings, it seems we can employ the split fermion idea \cite{splitfermions} to account for the fermion mass hierarchy from a set of domain wall couplings which are all about the same order of magnitude in this model. 

 Throughout the rest of the paper, we will quote the dimensionless background Yukawa couplings to five significant figures. The reasons for this are the exponential sensitivity of the profiles to these couplings and the difficulty that was found in generating the neutrino mass squared differences (which are quadratic in overlap integrals of these profiles) to an acceptable and reasonable precision. Since this is also a classical calculation where quantum corrections are ignored, and since the quark and neutrino masses are not as precisely measured or well known as those for the charged leptons, we will quote the resultant masses of the quarks and neutrinos to two significant figures, neutrino mass squared differences to one significant figure, and the charged lepton masses to three significant figures.

\subsection{The one-generation case with a Dirac neutrino and the suppression of colored-Higgs-induced proton decay}
\label{subsec:onegencase}

 In this section we shall show that the mass hierarchy amongst the first generation of fermions can be generated from the split fermion idea \cite{splitfermions} which arises naturally in our model. We will start with looking for solutions with the Higgs parameter choices of Eq. \ref{eq:firsthiggschoice}.

 Firstly, we must make the neutrino light. The right chiral neutrino is always localized at $y=0$ while the choice of Higgs parameters in Eq. \ref{eq:firsthiggschoice} (and in fact for those in Eqs. \ref{eq:secondhiggschoice} and \ref{eq:thirdhiggschoice} as well), the Higgs is localized to the right. Hence, the easiest way to induce a small Dirac neutrino mass is to shift the lepton doublet to the left. As the lepton doublet, $L$,  has hypercharge $-1$ and the charge conjugate of $d_R$ has hypercharge $+\frac{2}{3}$, choosing $\frac{\tilde{h_{5\chi}}}{\tilde{h_{5\eta}}}$ to be negative will displace the lepton doublet as desired while placing the right-chiral down quark to the right, near the electroweak Higgs.
\begin{figure}[h]
\begin{center}
\includegraphics[scale=0.7]{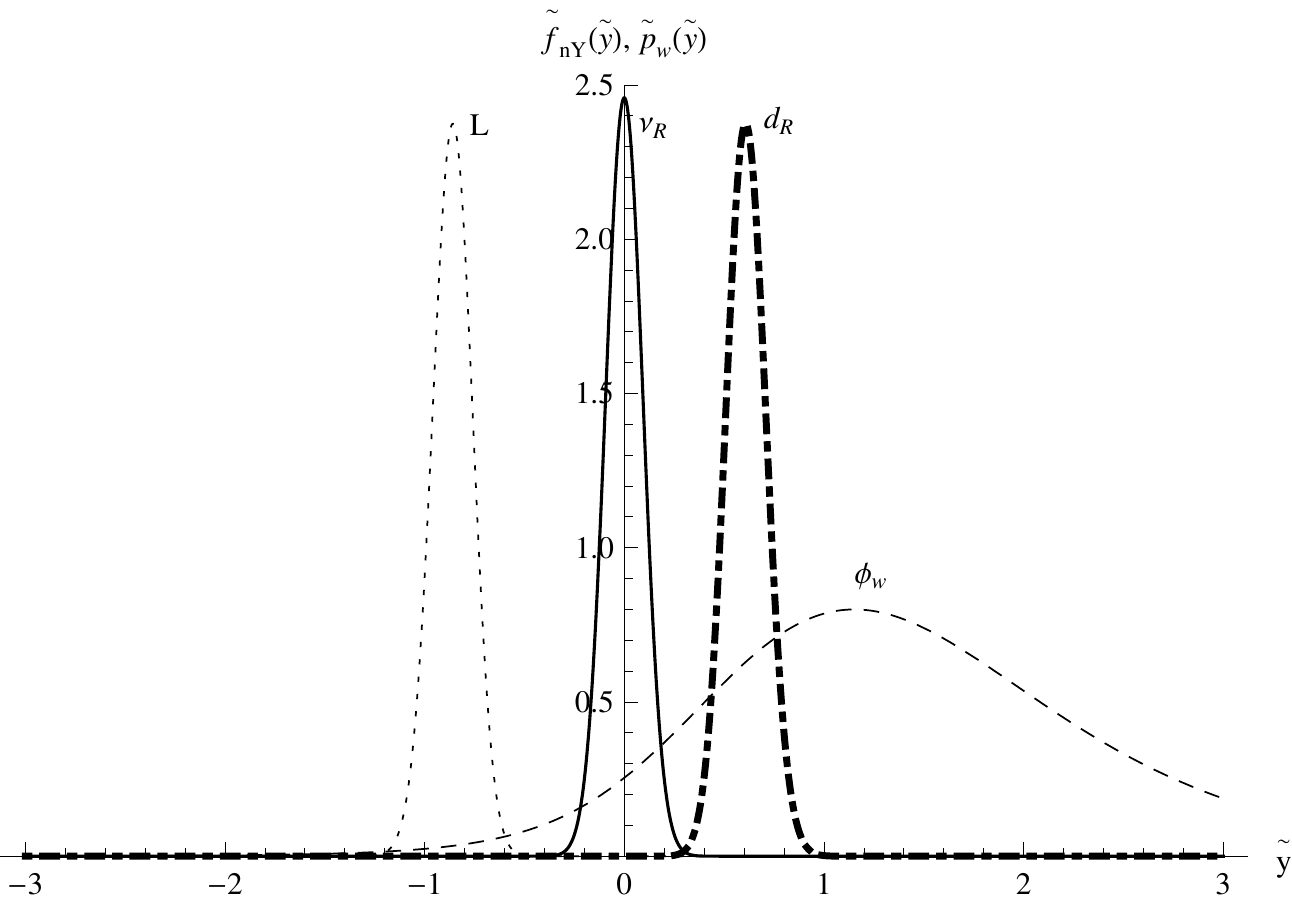}
\caption{The profiles for $\nu_R$, $L$, $d_R$ and the electroweak Higgs with the Higgs parameter choice of Eq. \ref{eq:firsthiggschoice}, $\tilde{h}_{1\eta}=115$, $\tilde{h}_{5\eta}=100$ and $\tilde{h}_{5\chi}=-250$.}
\end{center}
\end{figure}
We now need to make the charged fermion masses significantly larger. Since the charge conjugates of $u_R$ and $e_R$, and the quark doublet have hypercharges $-\frac{4}{3}$, $+2$ and $+\frac{1}{3}$ respectively, making the ratio $\frac{\tilde{h}_{10\chi}}{\tilde{h}_{10\eta}}$ positive will shift $e_R$ far to the left, towards the lepton doublet, $Q$ to slightly to the left, and $u_R$ to the right.
\begin{figure}[h]
\begin{center}
\includegraphics[scale=0.7]{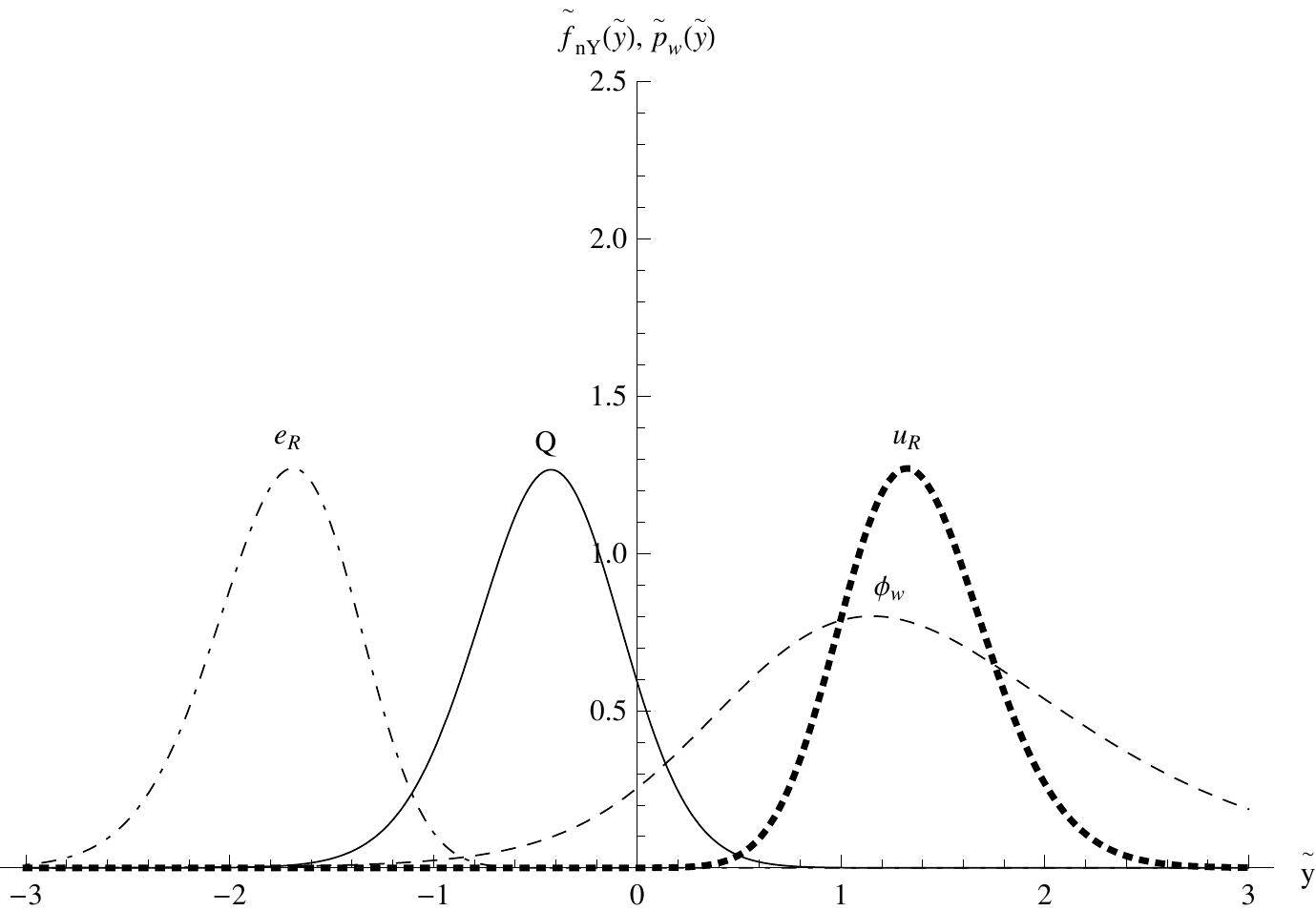}
\caption{The profiles for the 10 representation, and the electroweak Higgs with the Higgs parameter choice of Eq. \ref{eq:firsthiggschoice}, $\tilde{h}_{10\eta}=8.2674$ and $\tilde{h}_{10\chi}=27.911$.}
\end{center}
\end{figure}
We found the following solution by using this configuration, making the parameter choice $\tilde{h}_{5\eta}=100$ and $\tilde{h}_{5\chi}=-250$, plotting the contours along which the overlap integrals give the desired mass ratios,
\begin{figure}[h]
\begin{center}
\includegraphics[scale=0.7]{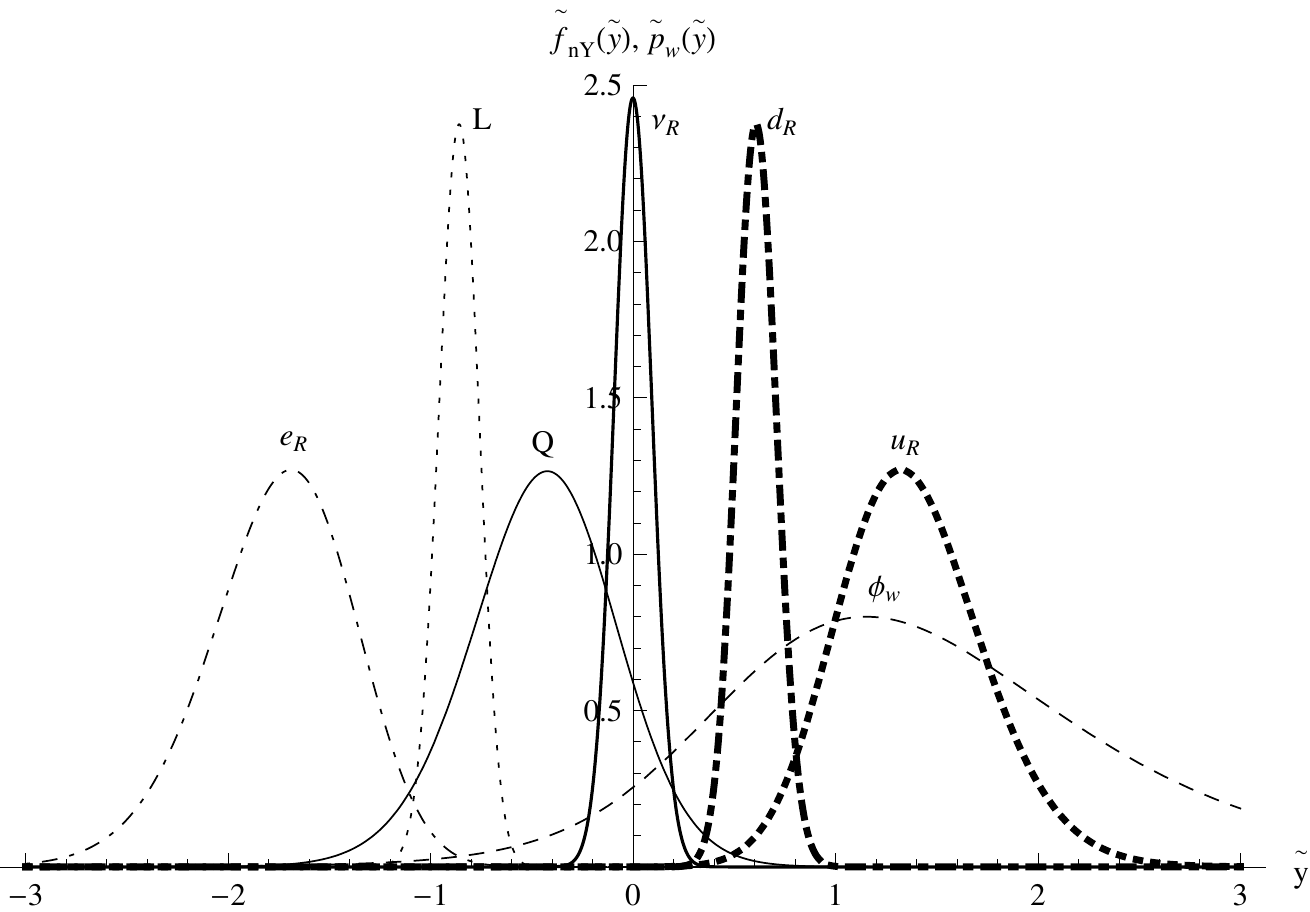}
\caption{The profiles for all fermions and $\phi_w$ for the first solution with $\tilde{h}_{1\eta}=115$, $\tilde{h}_{5\eta}=100$, $\tilde{h}_{5\chi}=-250$, $\tilde{h}_{10\eta}=8.2674$, $\tilde{h}_{10\chi}=27.911$ and the Higgs parameter choice of Eq. \ref{eq:firsthiggschoice}.}
\end{center}
\label{fig:firstonegensolprofiles}
\end{figure}
and then finding where the two contours intersected. Doing this yielded the solution for the couplings for the $10$ multiplet, $\tilde{h}_{10\eta}=8.2674$ and $\tilde{h}_{10\chi}=27.911$. With the ratios now fixed, setting the 5d electroweak Yukawas $h_{-}=h_{+}=h_{3}=5.2268\times{}10^{-3}k^{-\frac{1}{2}}$, and setting the kink coupling for the right handed neutrino to $\tilde{h}_{1\eta}=115$ gives the masses,
\begin{equation}
\begin{gathered}
m_{\nu}=0.13\;\rm{eV} \\
m_e = 0.511\;\rm{MeV} \\
m_u = 2.5\;\rm{MeV} \\
m_d = 5.0\;\rm{MeV.}
\end{gathered}
\label{eq:firstonegensolmasses}
\end{equation}

 Thus, we have generated an neutrino mass below the current most stringent upper bounds of roughly $2$ eV \cite{pdgquark} , the correct electron mass, and up and down quark masses within current constraints of $1.5$ MeV$<m_u<3.3$ MeV, and $3.5$ MeV$<m_d<6.0$ MeV \cite{pdgquark}.

\begin{figure}[h]
\subfloat[]{
\begin{fmffile}{protondecaygraph}
\fmfframe(15, 15)(15, 15){
\begin{fmfgraph*}(150, 75)
\fmfpen{thick}
\fmfstraight{}
\fmfleft{i1,i2,i3}
\fmfright{o1,o2,o3}
\fmf{fermion}{i1,v1}
\fmf{fermion}{o1,v1}
\fmf{fermion}{i2,v2}
\fmf{fermion}{o2,v2}
\fmf{fermion}{i3,o3}
\fmffreeze{}
\fmf{dashes, label=$\overline{\phi_c}$}{v1,v2}
\fmfdot{v1,v2}
\fmflabel{$u$}{i1}
\fmflabel{$u$}{i2}
\fmflabel{$d$}{i3}
\fmflabel{$e^+$}{o1}
\fmflabel{$\overline{d}$}{o2}
\fmflabel{$d$}{o3}
\fmflabel{$v_{ue}$}{v1}
\fmflabel{$v_{ud}$}{v2}
\end{fmfgraph*}
}
\end{fmffile}
}

\subfloat[]{
\begin{fmffile}{ptonupiplus}
\fmfframe(15, 15)(15, 15){
\begin{fmfgraph*}(150, 75)
\fmfpen{thick}
\fmfstraight{}
\fmfleft{i1,i2,i3}
\fmfright{o1,o2,o3}
\fmf{fermion}{i1,v1}
\fmf{fermion}{o1,v1}
\fmf{fermion}{i2,v2}
\fmf{fermion}{o2,v2}
\fmf{fermion}{i3,o3}
\fmffreeze{}
\fmf{dashes, label=$\overline{\phi_c}$}{v1,v2}
\fmfdot{v1,v2}
\fmflabel{$d$}{i1}
\fmflabel{$u$}{i2}
\fmflabel{$u$}{i3}
\fmflabel{$\nu_e$}{o1}
\fmflabel{$\overline{d}$}{o2}
\fmflabel{$u$}{o3}
\fmflabel{$v_{d\nu}$}{v1}
\fmflabel{$v_{ud}$}{v2}
\end{fmfgraph*}
}
\end{fmffile}
}
\caption{Feynman diagrams for the processes (a) $p\rightarrow{}e^+\pi^0$ and (b) $p\rightarrow{}\nu_e\pi^+$}
\label{fig:protondecayfeyndia}
\end{figure}
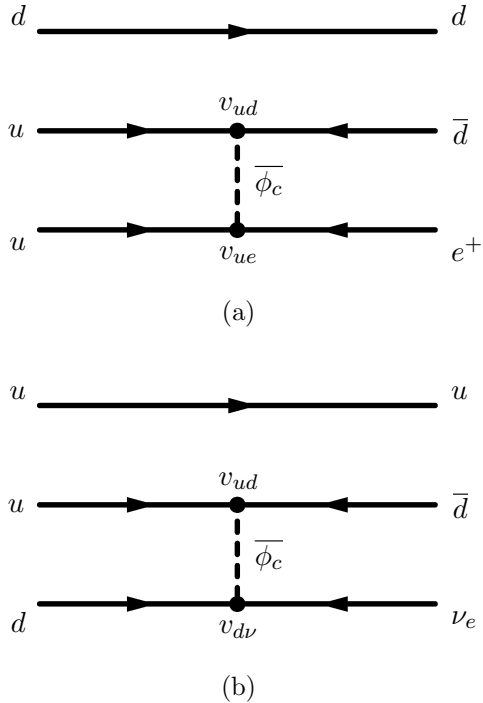

 Furthermore, it turns out we get significant suppression of some modes of colored Higgs-induced proton decay with this setup. The colored Higgs scalar can induce the decays $p\rightarrow{}e^+\pi^0$ and $p\rightarrow{}\nu_e\pi^+$, for which the Feynman diagrams are shown in Figures \ref{fig:protondecayfeyndia}(a) and \ref{fig:protondecayfeyndia}(b) respectively.

 For the process $p\rightarrow{}e^+\pi^0$, the partial lifetime of each contribution is 
\begin{equation}
\frac{m^4_c}{C^2_{v_{ue}}C^2_{v_{ud}}m^5_p},
\label{eq:protondecaypartiallifetime}
\end{equation}
where $C_{v_{ue}}$ and $C_{v_{ud}}$ are replaced by the effective 4d couplings strengths of the operators inducing the vertices $v_{ue}$ and $v_{ud}$ respectively. The operators responsible for the vertex $v_{ue}$ are $\overline{(e_R)^c}u_R\phi_c$ and $\overline{L}Q\phi_c$, and their respective coupling strengths are
\begin{equation}
\begin{gathered}
C_{\overline{(e_R)^c}u_R\phi_c} = 4h_{+}\int{}f_{u_R}(y)f_{e_R}(y)p_c(y)\;dy, \\
C_{\overline{L}Q\phi_c} = \frac{1}{\sqrt{2}}h_{-}\int{}f_{L}(y)f_{Q}(y)p_c(y)\;dy.
\label{eq:protondecv1couplings}
\end{gathered}
\end{equation}
 The operators responsible for the vertex $v_{ud}$ are $\overline{(u_R)^c}d_R\phi^*_c$ and $\epsilon^{ijk}Q_iQ_j(\phi^*_c)_k$, and the associated coupling strengths are 
\begin{equation}
\begin{gathered}
C_{\overline{(u_R)^c}d_R\phi^*_c} = \frac{1}{\sqrt{2}}h_{-}\int{}f_{u_R}(y)f_{d_R}(y)p_c(y)\;dy, \\
C_{QQ\phi^*_c} = 4h_{+}\int{}(f_{Q}(y))^2p_c(y)\;dy.
\end{gathered}
\label{eq:protondecv2couplings}
\end{equation} 

 Similarly, the partial lifetime of each contribution to $p\rightarrow{}\nu_e\pi^+$ is 
\begin{equation}
\frac{m^4_c}{C^2_{v_{d\nu}}C^2_{v_{ud}}m^5_p}.
\label{eq:ptonupipluslifetime}
\end{equation}
The operators responsible for the vertex $v_{ud}$ in the $p\rightarrow{}\nu_e\pi^+$ are the same as that for $p\rightarrow{}e^+\pi^0$, while the operators responsible for the $v_{d\nu}$ vertex are $\overline{L}Q\phi_c$ and $\overline{(\nu_R)^c}d_R\phi_c$. The coupling strength for the operator $\overline{(\nu_R)^c}d_R\phi_c$ is 
\begin{equation}
C_{\overline{(\nu_R)^c}d_R\phi_c} = h_{3}\int{}f_{\nu_R}(y)f_{d_R}(y)p_c(y)\;dy.
\label{eq:protondecvnucoupling}
\end{equation}

 For the solution given above, it turns out that the partial width for $p\rightarrow{}e^{+}\pi^{0}$ involving just right chiral fermions is substantially suppressed, with $C_{\overline{(e_R)^c}u_R\phi_c}=9.6\times{}10^{-14}$ and $C_{\overline{(u_R)^c}d_R\phi^*_c}=2.1\times{}10^{-5}$. Since the partial lifetime for $p\rightarrow{}e^{+}\pi^{0}$ is at least $8.2\times{}10^{33}$ years \cite{superkprotondecaylimit2009}, and given $m_p=0.938\;\rm{GeV}$ \cite{pdgquark}, this sets a lower bound on the colored Higgs mass of about $3.3\times{}10^4\;\rm{TeV}$, much reduced compared to the standard result of $m_c\sim\Lambda_{GUT}\sim10^{16}\;\rm{GeV}$. Since $C_{\overline{(\nu_R)^c}d_R\phi_c}=3.7\times{}10^{-8}$, and the lower bound of the partial lifetime for $p\rightarrow{}\nu\pi^+$ is $2.5\times{}10^{31}$ years \cite{pdgquark}, the contribution to $p\rightarrow{}\nu_e\pi^+$ involving the vertices $\overline{(\nu_R)^c}d_R\phi_c$ and $\overline{(u_R)^c}d_R\phi^*_c$ sets a lower bound on the colored Higgs mass of $4.8\times{}10^{6}\;\rm{TeV}$.

 However the contribution involving just left-chiral fermions is not substantially suppressed from the splittings, with $C_{\overline{L}Q\phi_c}=9.1\times{}10^{-4}$ and $C_{QQ\phi^*_c}=1.0\times{}10^{-2}$. These operators contribute to  both $p\rightarrow{}e^{+}\pi^{0}$ and $p\rightarrow{}\nu_e\pi^+$, and thus the partial widths coming from the combination of these operators set lower limits on the colored Higgs mass of $7.0\times{}10^{13}\;GeV$ for $p\rightarrow{}e^{+}\pi^{0}$, and $1.6\times{}10^{13}\;\rm{GeV}$ to suppress $p\rightarrow{}\nu_e\pi^+$. Therefore, we must still fine-tune so that the colored Higgs mass is of the order $\sim\;10^{14}$ GeV to suppress all proton decay modes induced by the colored Higgs scalar.

 One might then ask how to suppress proton decay even further. We could try looking for a solution where the profiles are more spread out in the extra dimension. It turns out the choice of parameters, $\tilde{h}_{1\eta}=100, \; \tilde{h}_{5\eta}=100, \; \tilde{h}_{5\chi}=-700, \; \tilde{h}_{10\eta}=0.81688, \; \tilde{h}_{10\chi}=23.868$, and $h_+=h_{-}=h_3=0.11177k^{-\frac{1}{2}}$, yields the same masses for the electron and the quarks as the first solution, and gives a neutrino mass of the order
\begin{equation}
m_{\nu}\sim10^{-24}\;\rm{eV.}
\label{eq:secondonegensolmasses}
\end{equation}
\begin{figure}[h]
\begin{center}
\includegraphics[scale=0.7]{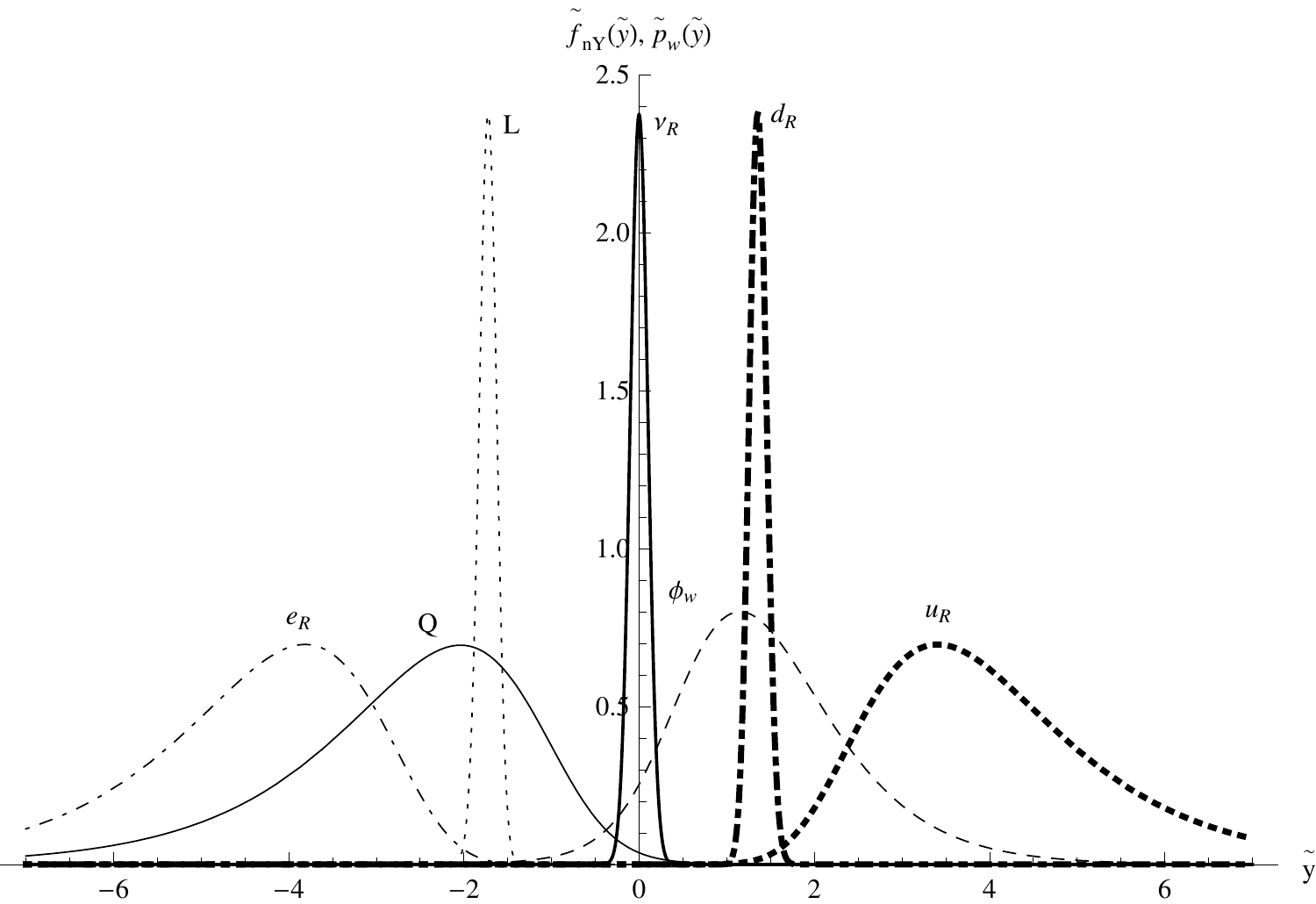}
\caption{The spread of profiles for the second solution with $\tilde{h}_{1\eta}=100$, $\tilde{h}_{5\eta}=100$, $\tilde{h}_{5\chi}=-700$, $\tilde{h}_{10\eta}=0.81688$, $\tilde{h}_{10\chi}=23.868$, and the Higgs parameter choice of Eq. \ref{eq:firsthiggschoice}.}
\end{center}
\label{fig:secondonegensolprofiles}
\end{figure}

 For proton decay, we now have for the operators involving just right chiral fermions $C_{\overline{(e_R)^c}u_R\phi_c}=4.2\times{}10^{-17}$, $C_{\overline{(u_R)^c}d_R\phi^*_c}=1.7\times{}10^{-5}$, and $C_{\overline{(\nu_R)^c}d_R\phi_c}=2.8\times{}10^{-23}$, which yield $m_c>6.1\times{}10^2$ TeV from the partial width for $p\rightarrow{}e^{+}\pi^{0}$, and $m_c>1.2\times{}10^2$ GeV from the partial width contributed to $p\rightarrow{}\nu_e\pi^+$. However the couplings for the operators involving the left chiral fermions are still not suppressed enough to solve the doublet-triplet splitting problem naturally, with $C_{\overline{L}Q\phi_c}=2.0\times{}10^{-2}$ and $C_{QQ\phi^*_c}=0.21$, setting the bound $m_c>1.5\times{}10^{15}$ GeV from the more constraining decay $p\rightarrow{}e^{+}\pi^{0}$.

 By spreading out the profiles, we have increased the spread of the domain wall parameters while only suppressing the proton decay modes induced from right chiral fermions by a further two orders of magnitude. It turns out that a choice of Higgs potential parameters giving more localized Higgs profiles can solve the first problem while yielding a similar result for proton decay. A solution for the second Higgs profile for which the Higgs parameters are those in Eq. \ref{eq:secondhiggschoice} is $h_+=h_{-}=h_3=82.975k^{-\frac{1}{2}}, \; \tilde{h}_{1\eta}=200, \; \tilde{h}_{5\eta}=100,\; \tilde{h}_{5\chi}=-250, \; \tilde{h}_{10\eta}=60.126, \; \tilde{h}_{10\chi}=99.829$, which again yields the same masses for the electron and the quarks as the previous solutions, and the neutrino mass
\begin{equation}
m_{\nu}=0.024 \;\rm{eV.} 
\label{eq:thirdonegensolmasses}
\end{equation}
 Interestingly, for these parameters, $C_{\overline{(e_R)^c}u_R\phi_c}=2.0\times{}10^{-24}$ and $C_{\overline{(u_R)^c}d_R\phi^*_c}=1.1\times{}10^{-3}$ suppressing the mode of $p\rightarrow{}e^{+}\pi^{0}$ involving just the right chiral fermions to the extent that the lower bound for the colored Higgs mass set by this mode is just $1.1$ TeV. However, the decays involving the left chiral fermions are in fact enhanced rather than suppressed by the fermion splittings with $C_{\overline{L}Q\phi_c}=2.0\times{}10^{-2}$ and $C_{QQ\phi^*_c}=39$, which means that the suppression factor coming from the effective coupling constants is of order 1, and so we need to make $m_c\sim{}M_{GUT}$. Also, the partial width involving the right chiral fermions for $p\rightarrow{}\nu_e\pi^+$ is not as suppressed this time, with $C_{\overline{(\nu_R)^c}d_R\phi_c}=8.5\times{}10^{-7}$, so that an $m_c$ of order $10^{11}\;\rm{GeV}$ is required to suppress this particular mode. 
\begin{figure}[h]
\begin{center}
\includegraphics[scale=0.7]{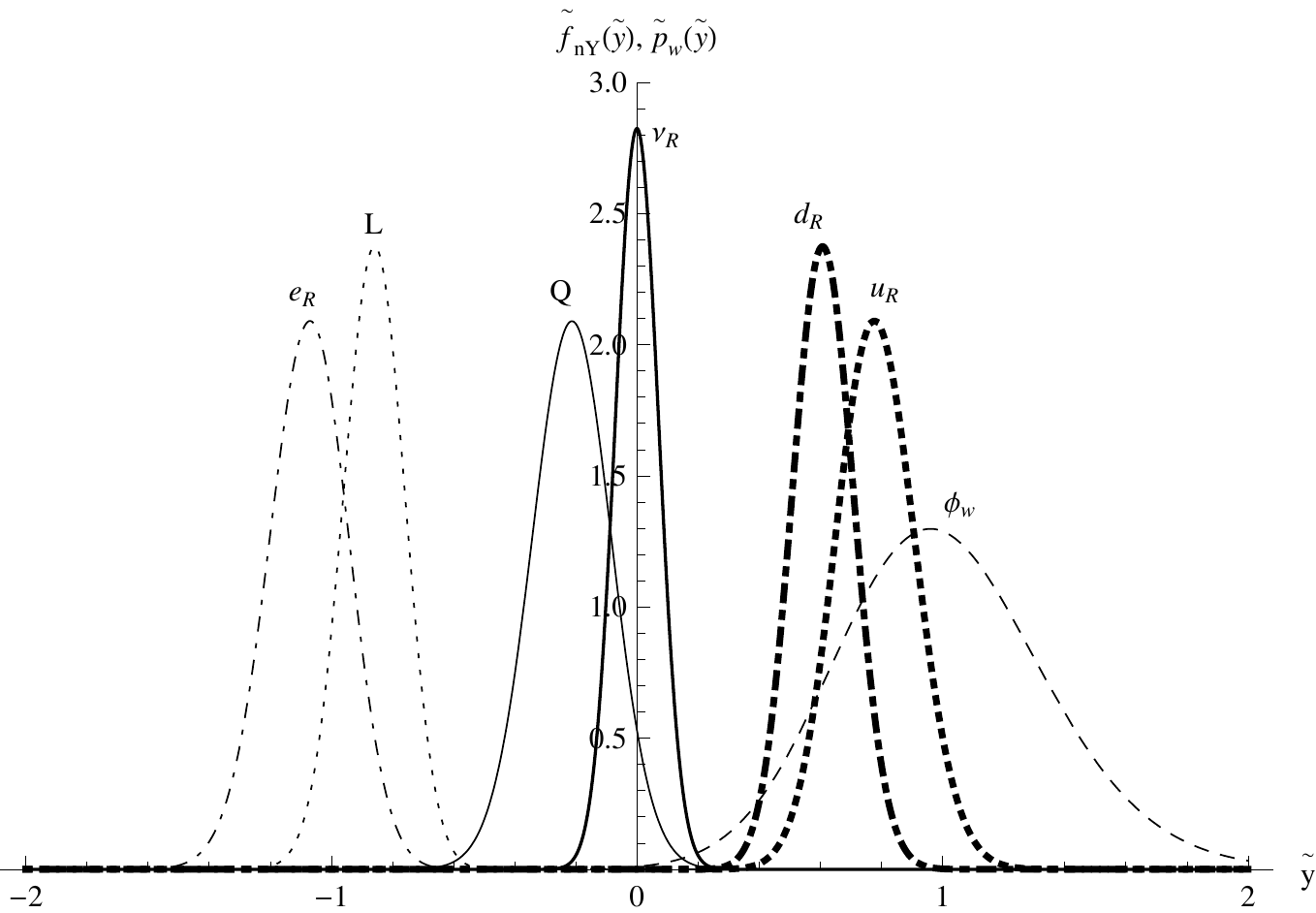}
\caption{The profiles for the solution with $\tilde{h}_{1\eta}=200$, $\tilde{h}_{5\eta}=100$, $\tilde{h}_{5\chi}=-250$, $\tilde{h}_{10\eta}=60.126$, $\tilde{h}_{10\chi}=99.829$, and the choice of Higgs parameters in Eq. \ref{eq:secondhiggschoice}.}
\end{center}
\label{fig:thirdonegensolprofiles}
\end{figure}

 The ultimate reason we have successfully suppressed the modes of proton decay involving just the right chiral fermions but not those involving the left chiral fermions so far was that the vertices involving the right chiral fermions depended on the profiles for $u_R$ and $d_R$ which were localized near the electroweak Higgs away from the colored Higgs, whereas due to the setup to generate the mass hierarchy, the quark and lepton doublets were placed significantly closer to the colored Higgs. To keep the natural solution to the mass hierarchy problem, we do not wish to displace the quark and lepton doublets; a more fruitful option is to choose Higgs parameters such that the colored Higgs is well displaced from the domain wall, while at the same time the electroweak Higgs is close to $y=0$. We have seen in Sec. \ref{sec:higgsloc} that this is in fact possible with the Higgs parameter choice given in Eq. \ref{eq:thirdhiggschoice}. A solution for this third Higgs profile to the mass hierarchy problem is the parameter choice $h_+=h_{-}=h_3=40987k^{-\frac{1}{2}}, \; \tilde{h}_{1\eta}=1000, \; \tilde{h}_{5\eta}=1000,\; \tilde{h}_{5\chi}=-1000, \; \tilde{h}_{10\eta}=624.62, \; \tilde{h}_{10\chi}=382.43$, which yields the same electron, up and down quark masses as before and a neutrino mass of the order
\begin{equation}
m_{\nu}\sim10^{-4} \;\rm{eV.} 
\label{eq:fourthonegensolmasses}
\end{equation}

 This time, for the proton decay inducing interactions, $C_{\overline{(e_R)^c}u_R\phi_c}\sim10^{-139}$, $C_{\overline{(u_R)^c}d_R\phi^*_c}\sim10^{-131}$ and $C_{\overline{(\nu_R)^c}d_R\phi_c}\sim10^{-126}$ for the operators involving just right chiral fermions and for those involving the left chiral fermions, $C_{\overline{L}Q\phi_c}\sim10^{-92}$ and $C_{QQ\phi^*_c}\sim10^{-98}$. Hence, all the decay modes are suppressed by roughly 90-100 orders of magnitude, with the most constraining decay mode $p\rightarrow{}e^{+}\pi^{0}$ with the left chiral fermions now setting a lower bound on the colored Higgs mass of $\sim{}10^{-69}$ eV. Realistically, for such a solution, the colored Higgs mass should still at the very least be $45$ GeV since we have not seen the Z boson decay into them, and more probably $\sim1$ TeV since it is proportional to $k$ in this model, so the partial lifetime arising from colored Higgs induced proton decay would be over $10^{100}$ years. 
\begin{figure}[h]
\begin{center}
\includegraphics[scale=0.7]{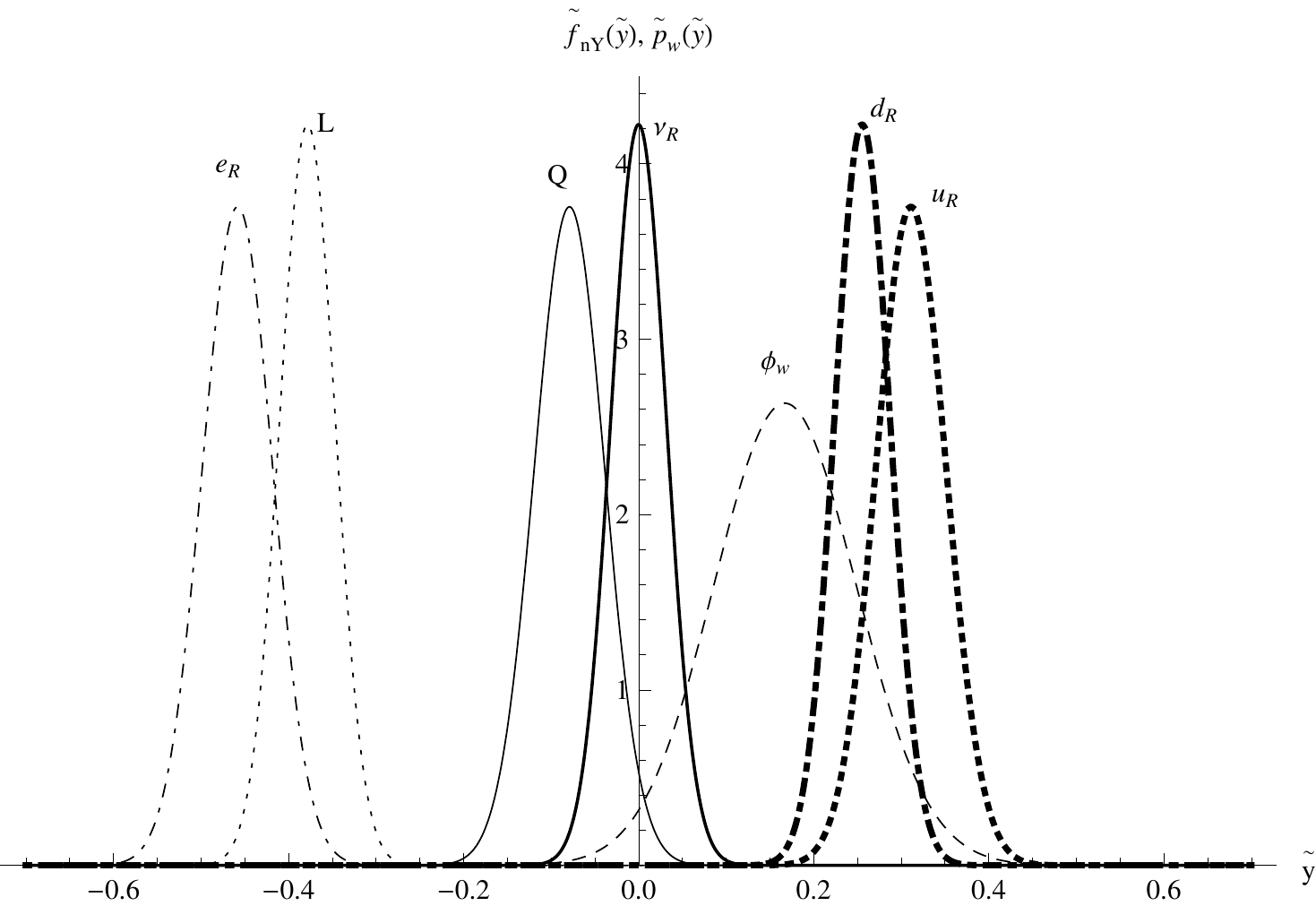}
\caption{The profiles for the solution with $\tilde{h}_{1\eta}=1000$, $\tilde{h}_{5\eta}=1000$,$\tilde{h}_{5\chi}=-1000$, $\tilde{h}_{10\eta}=624.62$, $\tilde{h}_{10\chi}=382.43$, and the choice of Higgs parameters in Eq. \ref{eq:thirdhiggschoice}.}
\end{center}
\label{fig:fourthonegensolprofiles}
\end{figure}

 The profiles for the latter two solutions are much less dispersed than those first two for the most delocalized electroweak Higgs profile, as one might expect. This is reflected in the breadth of the domain wall parameters; for the third and fourth solutions, the ratios of the magnitude of the largest background coupling ($\tilde{h}_{5\chi}=-250$ and $\tilde{h}_{1\eta}=\tilde{h}_{5\eta}=-\tilde{h}_{5\chi}=1000$ respectively) to the smallest ($\tilde{h}_{10\eta}=60.126$ and $\tilde{h}_{10\chi}=382.43$) are roughly $4.2$ and $2.6$ respectively, under an order of magnitude. In comparison, for the first two solutions, the corresponding ratios are about 30.2 and 860 respectively. This difference, as we will see, is exacerbated for three generations. 

 In summary, we have shown that the one generation mass hierarchy can be generated by splitting the fermions without fine tuning the electroweak Yukawa constants, and that for an appropriate choice of Higgs parameters, one can reduce the spread of the domain wall Yukawa constants and suppress proton decay by roughly 100 orders of magnitude without fine tuning the colored Higgs mass. In the next section, we will see that we can also do this for the three generation case without quark and lepton mixing.

\subsection{Generating the higher generation mass hierarchies without electroweak mixing}

 For the sake of simplicity, we can solve the mass hierarchy problem while first omitting quark and lepton mixing by setting the off diagonal elements of the electroweak Yukawa matrices to zero. Solutions are found in analogous fashion to the one generation case by finding where the overlap integrals are in the desired ratios in domain wall parameter space. 

 For the first Higgs profile, setting $h^i_{+}= h^i_{-} =h^i_3=1.4093k^{-\frac{1}{2}}$, the set of parameters choices for the domain wall Yukawa parameters and the resultant masses $m^i_E$ for the electron-type leptons, $m^i_U$ for the up-type quarks and $m^i_D$ for the down-type quarks are shown in Table \ref{tab:sol1}. As one can see, these masses all lie within current experimental limits \cite{pdgquark}. We then get similar results for the second Higgs profile with $h^i_{+}=h^i_{-}=h^i_3=7859.3k^{-\frac{1}{2}}$ and the third Higgs profile with $h^i_{+}=h^i_{-}=h^i_3=2701.2k^{-\frac{1}{2}}$, with the solutions for these two parameter choices given in Tables \ref{tab:sol2} and \ref{tab:sol3} respectively.

\begin{table}[h!]
\begin{center}
\begin{tabular}{|l|l|l|l|l|l|l|l|}
\hline
 & & & & & & & \\
$i$ & $\tilde{h}^i_{5\eta}$ & $\tilde{h}^i_{5\chi}$ & $\tilde{h}^i_{10\eta}$ & $\tilde{h}^i_{10\chi}$ & $m^i_{E}$(MeV) & $m^i_{U}$(MeV) & $m^i_{D}$(MeV) \\ \hline
1 & 1064.0 & -8563.9 & 0.2 & 25.496 & 0.511 & 2.5 & 5.0 \\ \hline
2 & 48.986 & -708.28 & 1.5 & 17.330 & 106 & 1.3$\times{}10^3$ & 1.0$\times{}10^2$ \\ \hline
3 & 100 & -300 & 10.537 & 8.6032 & 1.78$\times{}10^{3}$ & 1.7$\times{}10^{5}$ & 4.2$\times{}10^3$ \\
\hline
\end{tabular}
\end{center}
\caption{A set of domain wall parameters and the resultant masses with Higgs parameters chosen in Eq. \ref{eq:firsthiggschoice} and electroweak Yukawas set to $h^i_{+}= h^i_{-} =h^i_3=1.410k^{-\frac{1}{2}}$ for $i=1,2,3$}
\label{tab:sol1}
\end{table}
 
\begin{table}[h!]
\begin{center}
\begin{tabular}{|l|l|l|l|l|l|l|l|}
\hline
 & & & & & & & \\
$i$ & $\tilde{h}^i_{5\eta}$ & $\tilde{h}^i_{5\chi}$ & $\tilde{h}^i_{10\eta}$ & $\tilde{h}^i_{10\chi}$ & $m^i_{E}$(MeV) & $m^i_{U}$(MeV) & $m^i_{D}$(MeV) \\ \hline
1 & 200 & -648.41 & 38.552 & 99.220 & 0.511 & 2.5 & 5.0 \\ \hline
2 & 200 & -493.42 & 62.128 & 94.251 & 106 & 1.3$\times{}10^3$ & 1.0$\times{}10^2$ \\ \hline
3 & 200 & -400 & 73.744 & 76.383 & 1.78$\times{}10^3$ & 1.7$\times{}10^{5}$ & 4.2$\times{}10^3$\\
\hline
\end{tabular}
\end{center}
\caption{A set of domain wall parameters and the resultant masses with Higgs parameters chosen in Eq. \ref{eq:secondhiggschoice} and electroweak Yukawas set to $h^i_{+}=h^i_{-}=h^i_3=7859.3k^{-\frac{1}{2}}$ for $i=1,2,3$}
\label{tab:sol2}
\end{table}

\begin{table}[h!]
\begin{center}
\begin{tabular}{|l|l|l|l|l|l|l|l|}
\hline
 & & & & & & & \\
$i$ & $\tilde{h}^i_{5\eta}$ & $\tilde{h}^i_{5\chi}$ & $\tilde{h}^i_{10\eta}$ & $\tilde{h}^i_{10\chi}$ & $m^i_{E}$(MeV) & $m^i_{U}$(MeV) & $m^i_{D}$(MeV) \\ \hline
1 & 2000 & -1585.2 & 660.91 & 369.07 & 0.511 & 2.5 & 5.0 \\ \hline
2 & 2000 & -1434.5 & 744.05 & 325.26 & 106 & 1.3$\times{}10^2$ & 1.1$\times{}10^2$ \\ \hline
3 & 2000 & -1300 & 708.14 & 256.02 & 1.78$\times{}10^3$ & 1.7$\times{}10^{5}$ & 4.2$\times{}10^3$\\
\hline
\end{tabular}
\end{center}
\caption{A set of domain wall parameters and the resultant masses with Higgs parameters chosen in Eq. \ref{eq:thirdhiggschoice} and electroweak Yukawas set to $h^i_{+}=h^i_{-}=h^i_3=2701.2k^{-\frac{1}{2}}$ for $i=1,2,3$}
\label{tab:sol3}
\end{table}

 There exists a finite range of parameter space spanned by the remaining couplings $h_{1\eta}^i$ of the right handed neutrinos to the domain wall which fit the currently accepted squared neutrino mass differences of $\Delta{}m_{12}=7.9_{-0.5}^{+0.6}\times{}10^{-5}\;\rm{eV^2}$ \cite{kamland2005} and $\Delta{}m_{23}=2.74_{-0.26}^{+0.44}\times{}10^{-3}\;eV^2$ \cite{minos2006}, and cosmological constraints (in some models) on the sum of the masses $\sum{}m_{\nu}<0.3-0.6\;\rm{eV}$ \cite{hannestadsumneutrinomasses2006, bernadissumneutrinomasses2008}, for normal, inverted, and quasidegenerate neutrino mass hierarchies. Provided that the fermion doublets are sufficiently localized and displaced away from the domain wall at $y=0$, where the right handed neutrinos are always situated, and the electroweak Higgs, one can just adjust the couplings of the right handed neutrinos to the kink to get the desired masses and hierarchy. For each of the three solutions given in Tables \ref{tab:sol1}, \ref{tab:sol2} and \ref{tab:sol3}, three example parameter choices for the $\tilde{h}^i_{1\eta}$ yielding normal(N), quasidegenerate(Q), and inverted(I) neutrino mass hierarchies are given in Tables \ref{tab:nusol1}, \ref{tab:nusol2} and \ref{tab:nusol3} respectively.
\begin{table}[h!]
\begin{center}
\begin{tabular}{|l|l|l|l|l|l|l|}
\hline
Hierarchy & $\tilde{h}^1_{1\eta}$ & $\tilde{h}^2_{1\eta}$ & $\tilde{h}^3_{1\eta}$ & $m_{\nu_e}$(eV) & $m_{\nu_{\mu}}$(eV) & $m_{\nu_{\tau}}$(eV) \\ \hline
N & 100 & 15.44 & 110.4 & 1.5$\times{}10^{-41}$ & 0.0089 & 0.053  \\ \hline
Q & 18.919 & 13.764 & 106.61 & 0.10 & 0.10 & 0.086   \\ \hline
I & 19.503 & 14.219 & 300 & 0.051 & 0.052 & 7.4$\times{}10^{-8}$  \\ \hline
\end{tabular}
\begin{tabular}{|l|l|l|}
\hline
$\sum{}m_i$(eV) & $\Delta{}m_{21}^2$(eV$^2$) & $\Delta{}m_{32}^2$(eV$^2$) \\ \hline
0.062 & 8$\times{}10^{-5}$ & 3$\times{}10^{-3}$ \\ \hline
0.29 & 8$\times{}10^{-5}$ & -3$\times{}10^{-3}$ \\ \hline
0.10 & 8$\times{}10^{-5}$ & -3$\times{}10^{-3}$ \\ \hline
\end{tabular}
\end{center}
\caption{Solutions for normal, quasidegenerate and inverted neutrino mass hierarchies given the parameter choices given in Table \ref{tab:sol1}}
\label{tab:nusol1}
\end{table}
\begin{table}[h!]
\begin{center}
\begin{tabular}{|l|l|l|l|l|l|l|l|l|l|}
\hline
Hierarchy & $\tilde{h^1_{1\eta}}$ & $\tilde{h^2_{1\eta}}$ & $\tilde{h^3_{1\eta}}$ & $m_{\nu_e}$(eV) & $m_{\nu_{\mu}}$(eV) & $m_{\nu_{\tau}}$(eV)    \\ \hline
N & 200 & 132.73 & 262.60 & 3.5$\times{}10^{-15}$ & 0.0089 & 0.053    \\ \hline
Q & 54.564 & 114.28 & 253.20 & 0.096 & 0.096 & 0.081   \\ \hline
I & 56.690 & 118.95 & 650 & 0.051 & 0.052 & 2.9$\times{}10^{-6}$   \\ \hline
\end{tabular}
\begin{tabular}{|l|l|l|}
\hline
$\sum{}m_i$(eV) & $\Delta{}m_{21}^2$(eV$^2$) & $\Delta{}m_{32}^2$(eV$^2$) \\ \hline
0.062 & 8$\times{}10^{-5}$ & 3$\times{}10^{-3}$  \\ \hline
0.27 & 8$\times{}10^{-5}$ & -3$\times{}10^{-3}$   \\ \hline
0.10 & 8$\times{}10^{-5}$ & -3$\times{}10^{-3}$   \\ \hline
\end{tabular}
\end{center}
\caption{Solutions for normal, quasidegenerate and inverted neutrino mass hierarchies given the parameter choices given in Table \ref{tab:sol2}}
\label{tab:nusol2}
\end{table}

\begin{table}[h!]
\begin{center}
\begin{tabular}{|l|l|l|l|l|l|l|l|l|l|}
\hline
Hierarchy & $\tilde{h^1_{1\eta}}$ & $\tilde{h^2_{1\eta}}$ & $\tilde{h^3_{1\eta}}$ & $m_{\nu_e}$(eV) & $m_{\nu_{\mu}}$(eV) & $m_{\nu_{\tau}}$(eV)    \\ \hline
N & 2000 & 1449.2 & 2044.3 & 8.1$\times{}10^{-9}$ & 0.0089 & 0.053    \\ \hline
Q & 826.28 & 1250 & 1948.5 & 0.084 & 0.084 & 0.099   \\ \hline
I & 852.44 & 1291 & 4500 & 0.051 & 0.052 & 2.3$\times{}10^{-6}$   \\ \hline
\end{tabular}
\begin{tabular}{|l|l|l|}
\hline
$\sum{}m_i$(eV) & $\Delta{}m_{21}^2$(eV$^2$) & $\Delta{}m_{32}^2$(eV$^2$) \\ \hline
0.062 & 8$\times{}10^{-5}$ & 3$\times{}10^{-3}$  \\ \hline
0.27 & 8$\times{}10^{-5}$ & 3$\times{}10^{-3}$   \\ \hline
0.10 & 8$\times{}10^{-5}$ & -3$\times{}10^{-3}$   \\ \hline
\end{tabular}
\end{center}
\caption{Solutions for normal, quasidegenerate and inverted neutrino mass hierarchies given the parameter choices given in Table \ref{tab:sol3}}
\label{tab:nusol3}
\end{table}

\begin{figure}[H]
\begin{center}
\subfloat[]{\includegraphics[scale=0.80]{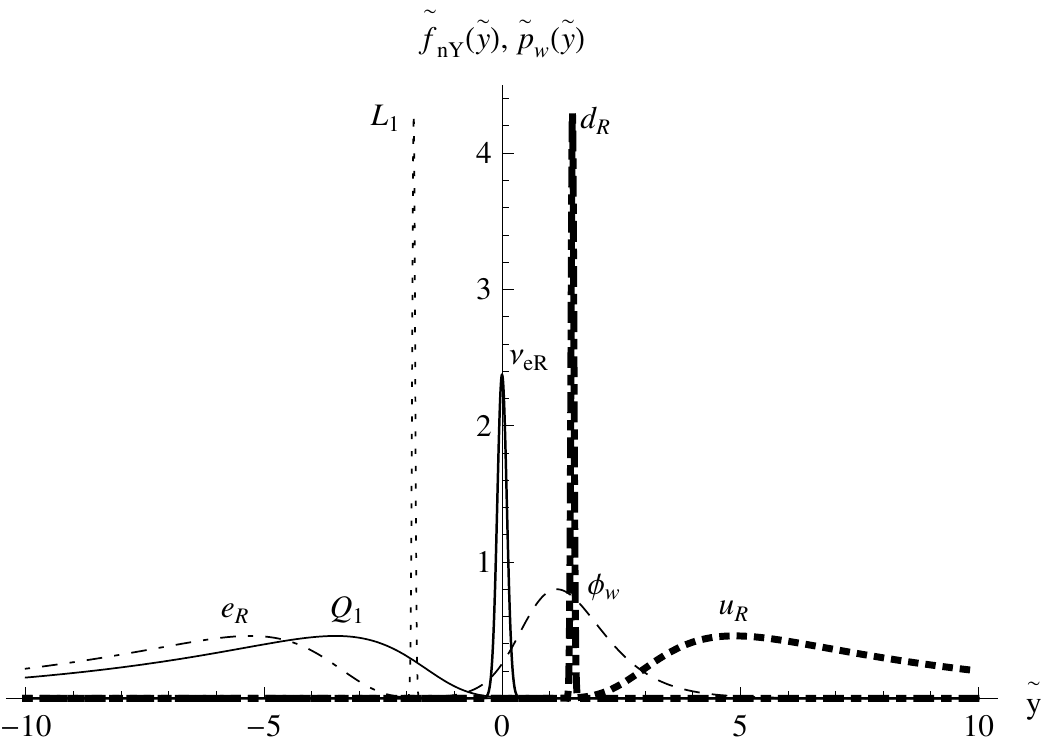}} \\
\subfloat[]{\includegraphics[scale=0.80]{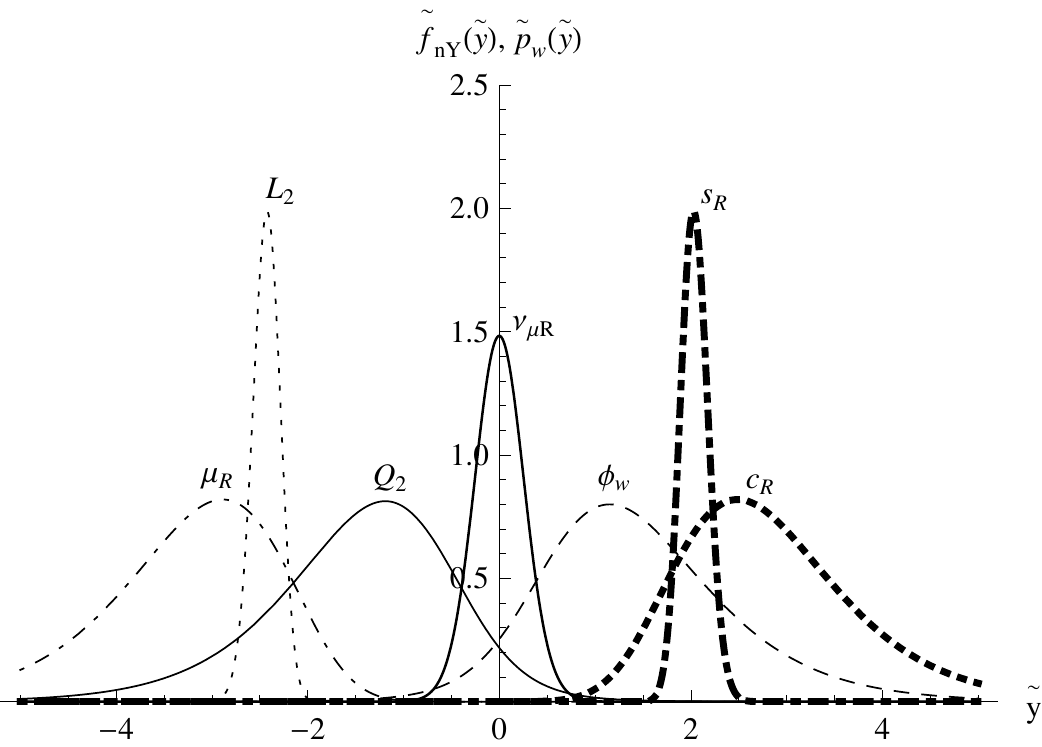}} \\
\subfloat[]{\includegraphics[scale=0.80]{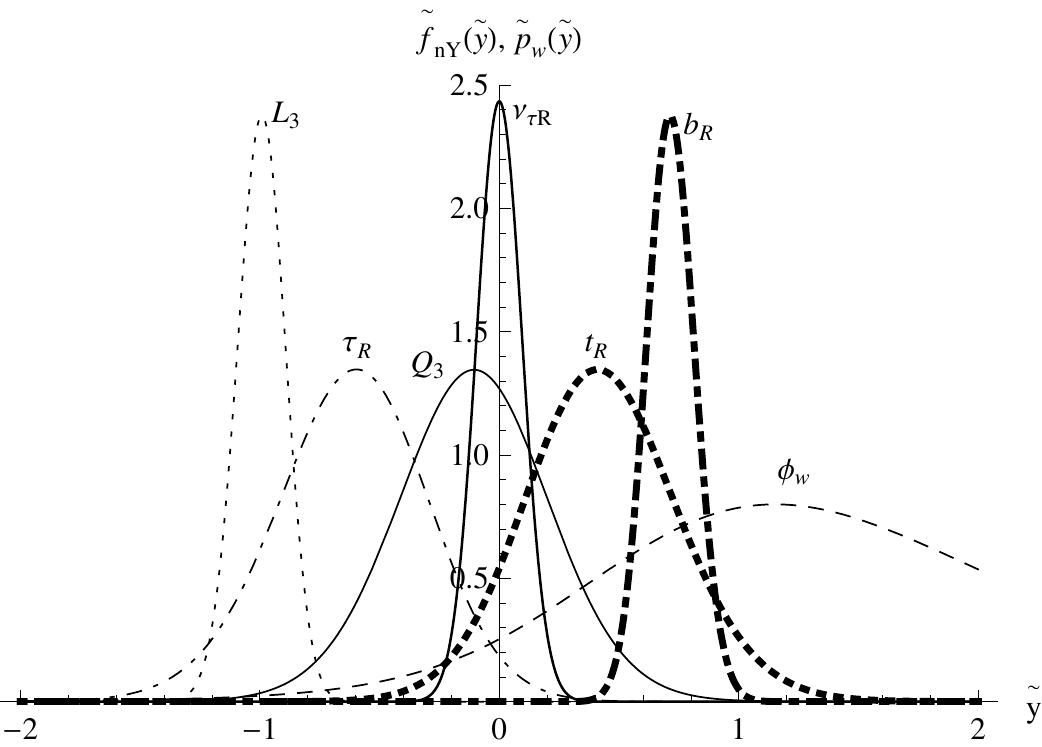}} \\
\caption{Plots of the profiles of the first (a), second (b), and third generation (c) of fermions with the parameter choice of Table \ref{tab:sol1} and the of the normal hierarchy parameter choice in Table \ref{tab:nusol1}}
\label{fig:sol1graphs}
\end{center}
\end{figure}

\begin{figure}[H]
\begin{center}
\subfloat[]{\includegraphics[scale=0.80]{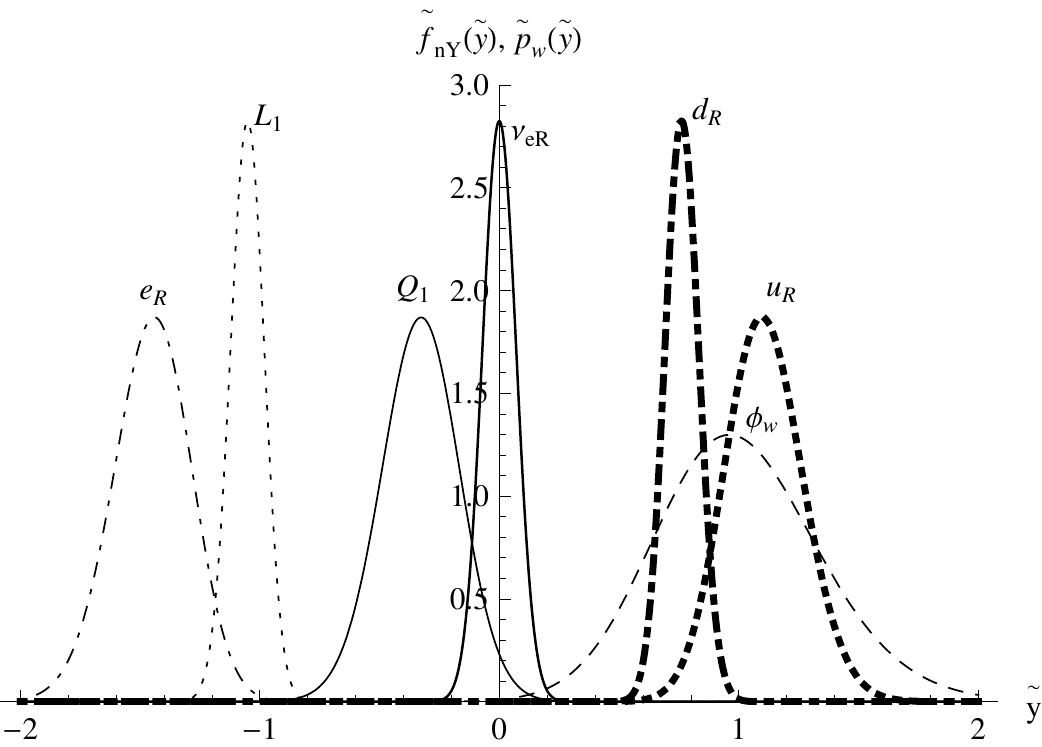}} \\
\subfloat[]{\includegraphics[scale=0.80]{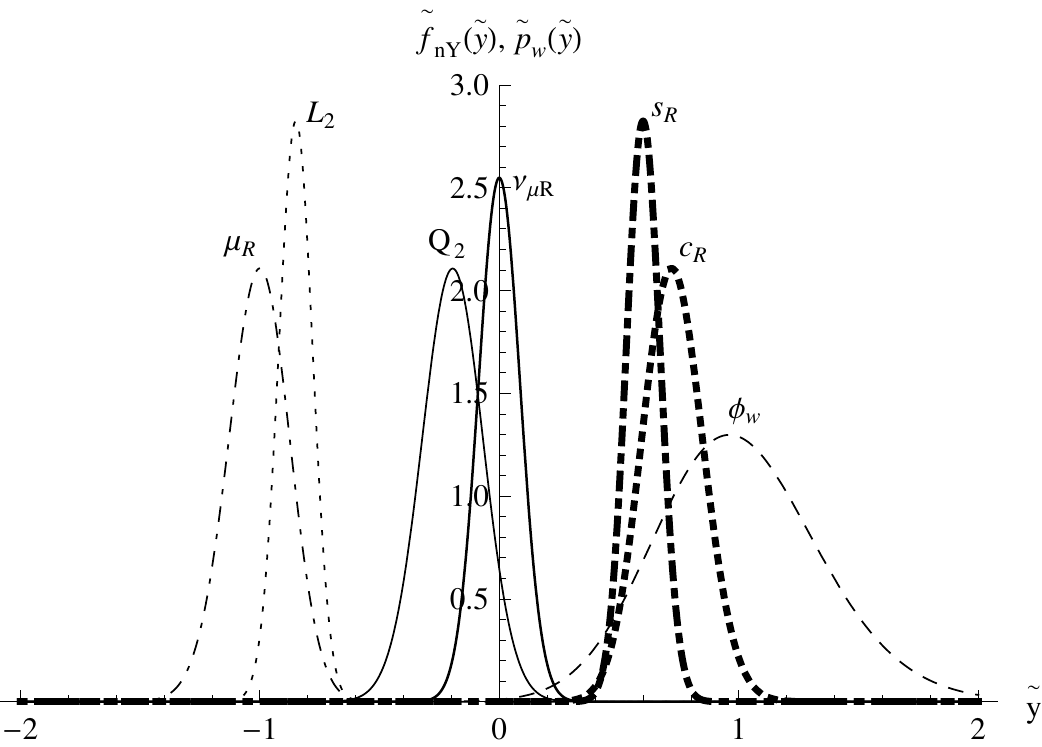}} \\
\subfloat[]{\includegraphics[scale=0.80]{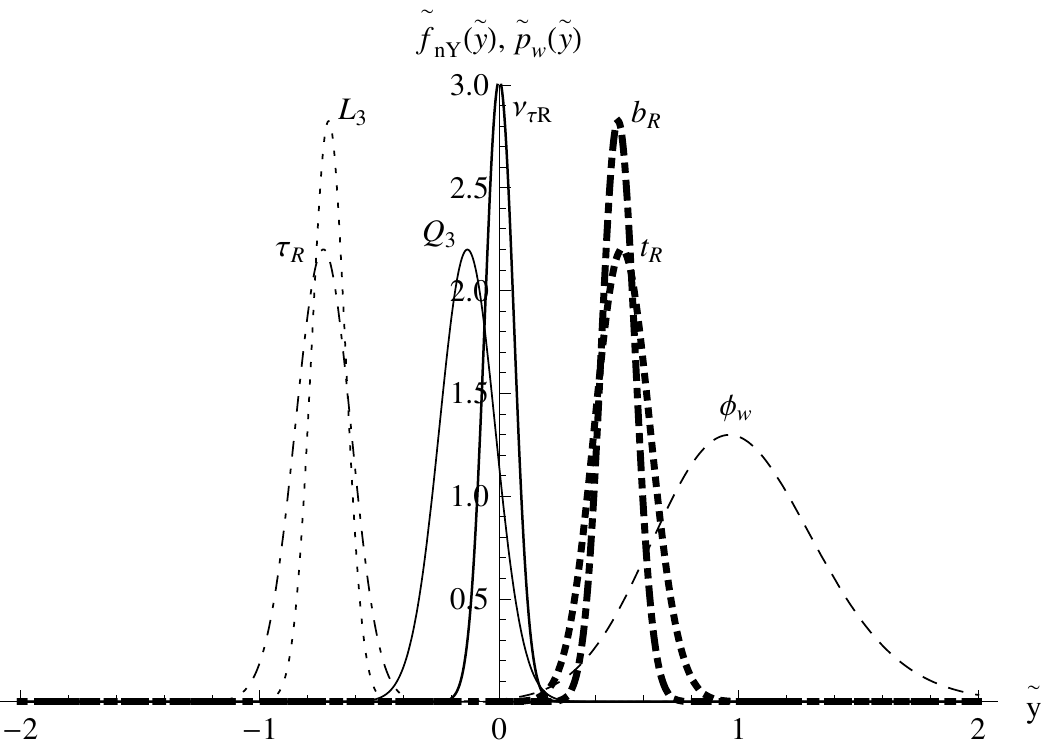}} \\
\caption{Plots of the profiles of the first (a), second (b), and third generation (c) of fermions with the parameter choice of Table \ref{tab:sol2} and the of the normal neutrino mass hierarchy parameter choice in Table \ref{tab:nusol2}}
\label{fig:sol2graphs}
\end{center}
\end{figure}

\begin{figure}[H]
\begin{center}
\subfloat[]{\includegraphics[scale=0.80]{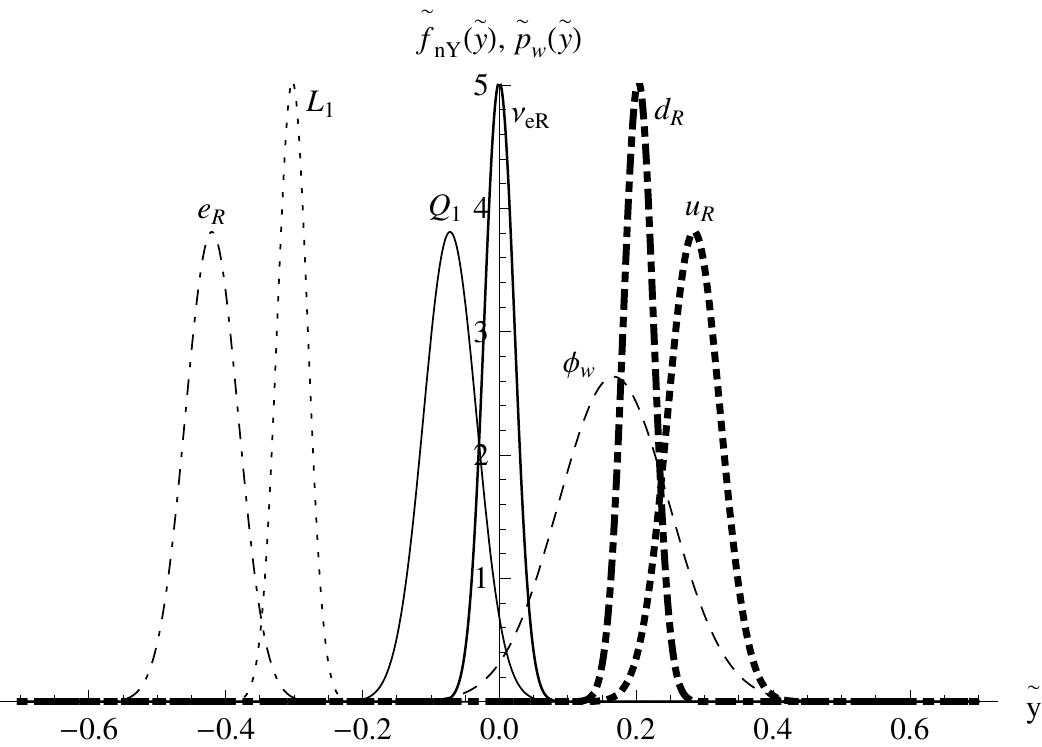}} \\
\subfloat[]{\includegraphics[scale=0.80]{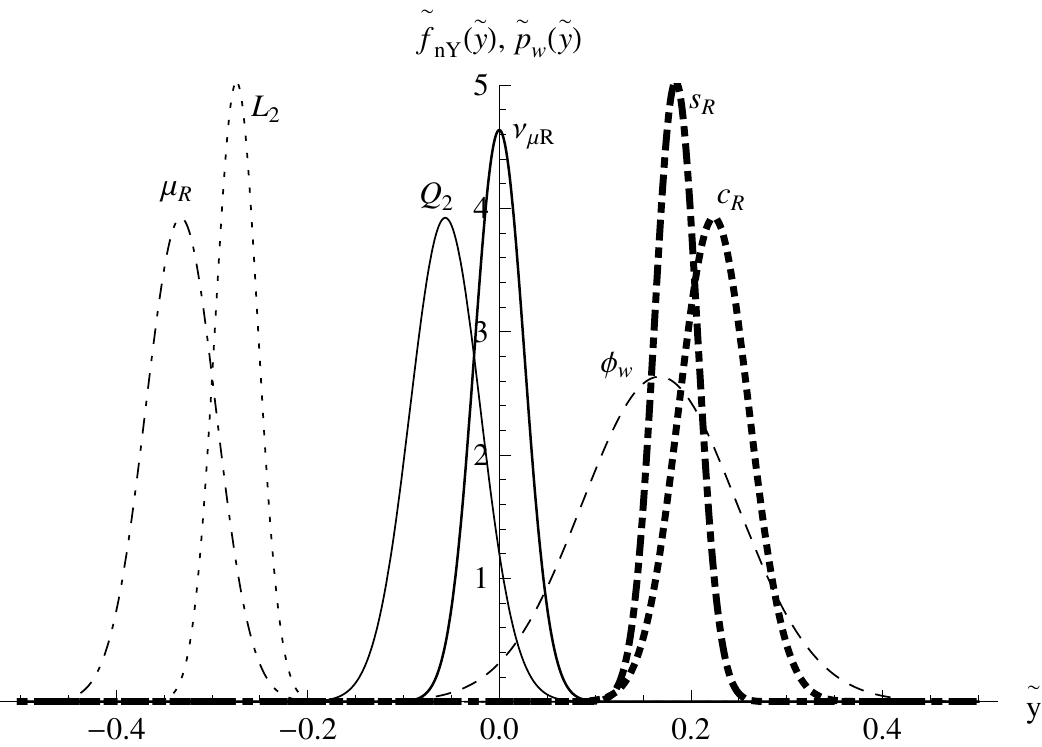}} \\
\subfloat[]{\includegraphics[scale=0.80]{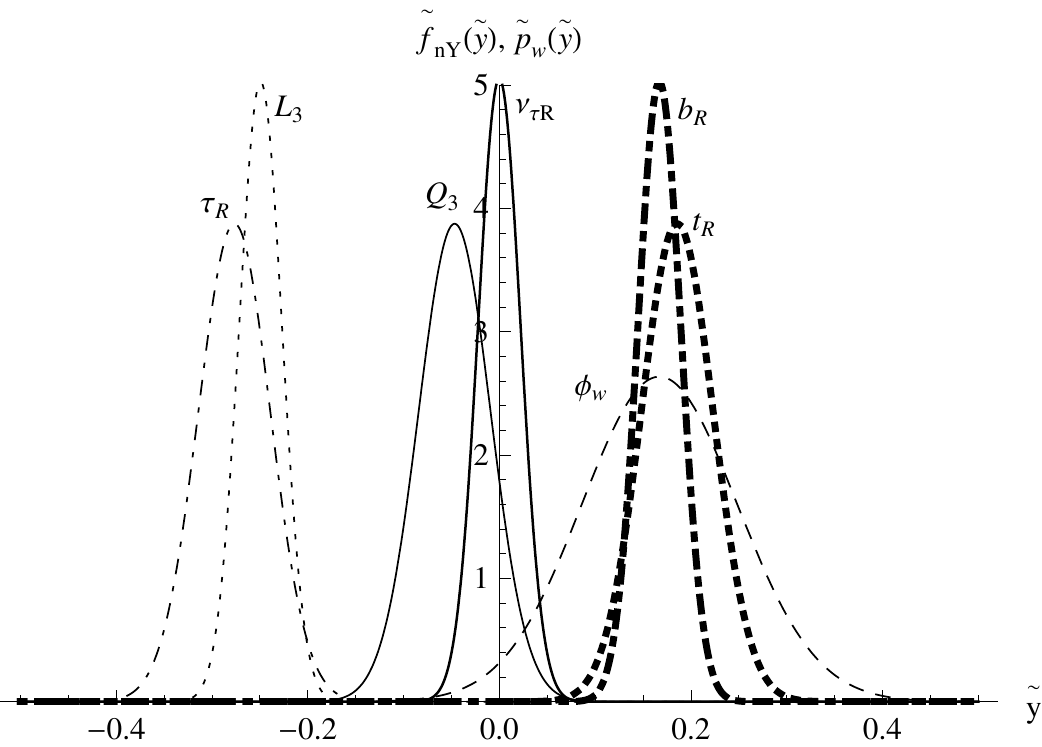}} \\
\caption{Plots of the profiles of the first (a), second (b), and third generation (c) of fermions with the parameter choice of Table \ref{tab:sol3} and the normal hierarchy parameter choice in Table \ref{tab:nusol3}}
\label{fig:sol3graphs}
\end{center}
\end{figure}

 The distribution of the fermions for each family for both the solutions with a normal neutrino mass hierarchy are shown in the Figs. \ref{fig:sol1graphs}, \ref{fig:sol2graphs} and \ref{fig:sol3graphs}. As can be seen, the lighter generations are, on average, more spread apart, more distant from $y=0$ and more delocalized. Comparing the plots for the first Higgs profile in Fig. \ref{fig:sol1graphs}  to those for the second and third Higgs profiles in Figs. \ref{fig:sol2graphs} and \ref{fig:sol3graphs}, it is conspicuous that this increase in spread of the fermions between generations is dramatically reduced for the more localized Higgs. This is reflected in the spread of domain wall parameters, with the ratios between the smallest and largest non-dimensionalized domain wall parameters. For the parameter choices of Tables \ref{tab:sol1}, \ref{tab:sol2}, and \ref{tab:sol3} and normal neutrino mass heirarchies, these ratios are respectively $4.3\times{}10^4$, $17$, and $7.8$. The solution of Table \ref{tab:sol3} is particularly interesting since the non-dimensionalized electroweak Yukawa constant $\tilde{h}_{-,+,3}^i=h_{-,+,3}^ik^{\frac{1}{2}}=2701.2$ is of the same order as the non-dimensionalized domain wall parameters.

 With regards to proton decay, the results for the parameter choices of Tables \ref{tab:sol1}, \ref{tab:sol2} and \ref{tab:sol3} are similar to those of Sec. \ref{subsec:onegencase}. For the first two solutions with Higgs parameters chosen from Eqs. \ref{eq:firsthiggschoice} and \ref{eq:secondhiggschoice}, the decay modes involving just right-chiral fermions are substantially suppressed while there is negligible suppression for the modes involving the left-chiral fermions. For the parameters chosen in Table \ref{tab:sol1}, $C_{\overline{(e_R)^c}u_R\phi_c}\sim10^{-18}$, $C_{\overline{(u_R)^c}d_R\phi^*_c}=1.7\times{}10^{-5}$, $C_{\overline{L}Q\phi_c}=5.7\times{}10^{-2}$ and $C_{QQ\phi^*_c}=0.80$, and hence the lower bound on the colored Higgs mass from the decay mode for $p\rightarrow{}e^+\pi^0$ involving just right-chiral fermions is roughly 120 TeV, while that from the decay mode involving just left-chiral fermions is of order $10^{15}$ GeV. 

 For the solution in Table \ref{tab:sol2}, $C_{\overline{(e_R)^c}u_R\phi_c}\sim10^{-30}$, $C_{\overline{(u_R)^c}d_R\phi^*_c}=4.1\times{}10^{-3}$, $C_{\overline{L}Q\phi_c}=2.5$ and $C_{QQ\phi^*_c}=7.4\times{}10^3$, so that the decay mode involving the right-chiral fermions sets a lower bound on $m_c$ of just 1.7 GeV, while the decay mode involving just left-chiral fermions are in fact enhanced, with the lower bound on $m_c$ increased to order $10^{17}$ GeV. 

 For the solution of Table \ref{tab:sol3}, just as with the one generation solution with the Higgs parameter choices of Eq. \ref{eq:thirdhiggschoice}, all decay modes are suppressed since the colored Higgs is well away from the domain wall and the electroweak Higgs. For this solution, we have $C_{\overline{(e_R)^c}u_R\phi_c}\sim10^{-135}$, $C_{\overline{(u_R)^c}d_R\phi^*_c}\sim10^{-129}$, $C_{\overline{L}Q\phi_c}\sim10^{-94}$ and $C_{QQ\phi^*_c}\sim10^{-99}$, so that the bound on $m_c$ set by the less suppressed decay mode involving the left-chiral fermions is of the order of $10^{-71}$ eV. For all neutrino mass hierachies, the coupling constant $C_{\overline{(\nu_R)^c}d_R\phi_c}$ is also well below $10^{-100}$, so that $p\rightarrow{}\nu{}_e\pi^+$ is also negligible.

 Now that it has been demonstrated that the three generation mass hierarchies can be generated from the exponential dependences of the overlaps on the domain wall couplings while suppressing proton decay, the next step is to incorporate quark and lepton mixing.

\subsection{Accounting for the Cabibbo angle in the two-generation case}

 To produce realistic mass matrices, we must account for the fermion mixing angles as well as the masses. For the sake of simplicity, we will work with two generations and show that the Cabibbo mixing angle can be produced along with the mass hierarchy.

 Performing the required analysis is quite complicated since, if we are to assume that all the 5d Yukawa couplings are equal, including the off diagonal couplings, then the order of the analogous equations giving desired mass matrix element ratios from the overlaps is the equal to the number of families in the theory. Thus, to generate the Cabibbo angle, one must solve equations which are quadratic in the overlaps, and which are also no longer separated with respect to the domain wall parameters. For the CKM matrix it is even worse since the equations are cubic. This raises difficulties, in particular, with the down and electron sectors, since these sectors depend on all of the background couplings of the charged fermions, which amount to eight for two generations. Hence, we are forced to start with the up quark sector first, for convenience, since it only depends on four couplings. This makes it difficult to guarantee that the Dirac neutrino masses will be light. 

 Instead of directly solving the equations quadratic in the overlaps, we will try to generate mass matrices approximately equal to the Cholesky decompositions of the desired mass matrices squared, $M^{\dagger}M$. This is similar to an approach of generating mass matrices in NNI (Nearest-Neighbour-Interaction) basis for the three generation case, as was done in the analysis with Gaussian profiles done in \cite{brancogouvea1}, and by Mirabelli and Schmaltz \cite{mirabellischmaltz2000}. The advantage of this approach is that we can now do the analysis in terms of equations linear in the overlaps instead. The main disadvantage is that we must rely on one of the off diagonal terms being significantly suppressed compared to the other couplings.

 It turns out that this approach can get the charged fermion mass hierarchies and the Cabbibo angle. Taking all the electroweak Yukawa couplings to be $h^{ij}_+=h^{ij}_{-}=h^{ij}_{3}=0.089104k^{-\frac{1}{2}}$, the Higgs parameters to be those of Eq. \ref{eq:firsthiggschoice}, and making the choices for the domain wall Yukawas in Table \ref{tab:cabsol},
\begin{table}[h!]
\begin{center}
\begin{tabular}{|l|l|l|l|l|}
\hline
  &  &  & &  \\
$i$ & $\tilde{h}^i_{5\eta}$ & $\tilde{h}^i_{5\chi}$ & $\tilde{h}^i_{10\eta}$ & $\tilde{h}^i_{10\chi}$\\ \hline
1 & 12.585 & -36.719 & 100 & 53.346 \\ \hline
2 & 365.78 & -1708.2 & 2.5273 & 27.095 \\ 
\hline
\end{tabular}
\end{center}\caption{The choices for the domain wall parameters with Higgs parameters from Eq. \ref{eq:firsthiggschoice} and $h^{ij}_+=h^{ij}_{-}=h^{ij}_{3}=0.089104k^{-\frac{1}{2}}$ for $i,j=1,2,3$}
\label{tab:cabsol}
\end{table}
we obtain the following mass matrices correct to three significant figures,

\begin{equation}
M_u=\begin{pmatrix}
1.04\times{}10^3 & 7.28\times{}10^2 \\
2.77\times{}10^{-3} & 3.05 \\
\end{pmatrix},
\end{equation}
\begin{equation}
M_d=\begin{pmatrix}
70.3 & 77.8 \\
\sim0 & 7.47 \\
\end{pmatrix},
\end{equation}
and 
\begin{equation}
M_e=\begin{pmatrix}
106 & \sim0 \\
0.465 & 0.511 \\
\end{pmatrix}.
\end{equation}
The resultant left diagonalization angle for $M_u$ is $\Theta_u=55^\circ$, and the left diagonalization angle for $M_d$ is $\Theta_d=42^\circ$, yielding the Cabibbo angle $\Theta_c=\Theta_u-\Theta_d=13^\circ$. Taking the square roots of the eigenvalues of $M^\dagger_uM_u$, $M^{\dagger}_dM_d$, and $M^\dagger_eM_e$ yields the masses of the charged fermions, and they turn out to be those in Table \ref{tab:cabsolmasses}.

\begin{table}[h!]
\begin{center}
\begin{tabular}{|l|l|l|l|}
\hline
$i$ & $m^i_{E}$(eV) & $m^i_{U}$(eV) & $m^i_{D}$(eV) \\ \hline
1 & 0.511 & 2.5 & 5.0 \\ \hline
2 & 106 & 1.3$\times{}10^3$ & 1.1$\times{}10^2$ \\ \hline
\end{tabular}
\end{center}
\caption{The masses of the mass eigenstates in the electron, up and down type sectors respectively with the parameter choice of Table \ref{tab:cabsol}}
\label{tab:cabsolmasses}
\end{table}

 This solution does not permit two light neutrino masses for the case of a Dirac neutrino. For example, for the parameter choice $\tilde{h}^1_{1\eta}=\tilde{h}^2_{1\eta}=100$, the mass matrix for the neutrino is 
\begin{equation}
M_{\nu}=\begin{pmatrix}
6.23 & \sim0 \\
6.23 & \sim0 \\
\end{pmatrix},
\end{equation}
which yields $m_{\nu{}1}=0$ and $m_{\nu{}2}=8.8\;MeV$. Because for this solution one of the left weak eigenstates of the neutrino has a $\tilde{h}^1_{5\chi}/\tilde{h}^1_{5\eta}$ ratio larger in magnitude, and thus is more distant from the wall and the Higgs, the mass matrix for the neutrino for this solution will always have the entries of one column being larger than the other for significantly large $\tilde{h}^i_{1\eta}$ and since one of the eigenvalues is typically of the same order as the larger of the two elements in the larger column in such a matrix. In the limit $\tilde{h}^i_{1\eta}\rightarrow \infty{}$ for $i=1,2$, 
\begin{equation}
\begin{split}
M^{ij}_u &=h_{3}\bar{v}\int{}f_{\nu^i_R}(y)f_{L^j}(y)p_w(y)\;dy \\
        &\rightarrow h_{3}\bar{v}f_{L^j}(0)p_w(0)
\end{split}
\end{equation}
as the profiles for the right handed neutrinos converge to delta functions at $y=0$, one obtains the mass matrix, 
\begin{equation}
M_{\nu}=\begin{pmatrix}
5.81 & \sim0 \\
5.81 & \sim0 \\
\end{pmatrix},
\end{equation}
which yields neutrino masses of $m_1=8.2\;\rm{MeV}$ and $m_2=0$. Thus, the best we can do for this solution which yields Cabibbo mixing and the charged mass hierarchy, for the case of a Dirac neutrino, is to generate a massless neutrino, and a neutrino about 2-3 MeV heavier than a down quark.

  This does not prove that there is no solution for a Dirac neutrino which incorporates quark mixing and the fermion mass hierarchy. The scheme we used and the section of parameter space searched led to one of the lepton doublets being placed too close to the right handed neutrino and too delocalized to support two light neutrinos. A more thorough search of the parameter space, perhaps utilizing a Monte Carlo method, will have to be done to determine whether a solution supporting two sufficiently light neutrinos exists for the case of a Dirac neutrino.

 For a Majorana neutrino, however, this solution presents no such problems. As was shown earlier in this paper, the seesaw mechanism can be employed in the model, and can thus be used to suppress the mass of the heavier neutrino.

 The set of domain wall parameters in Table \ref{tab:cabsol} generate the desired mass spectrum and the Cabibbo angle. The ratio between the parameters smallest ($\tilde{h}^2_{10\eta}=2.5273$) and largest ($\tilde{h}^2_{5\chi}=-1708.2$) in magnitude is roughly 670. To reduce this, we would need to find solutions with a more localized electroweak Higgs, such as those resulting from the parameter choices Eqs. \ref{eq:secondhiggschoice} and \ref{eq:thirdhiggschoice}. Finding such solutions has been difficult, however, and we will leave this to be done in a Monte Carlo search.

 Proton decay for this solution cannot be suppressed without fine-tuning the colored Higgs mass. Since we are using the Higgs parameter choice of Eq. \ref{eq:firsthiggschoice} and not that of Eq. \ref{eq:thirdhiggschoice}, the colored Higgs is sufficiently close to the domain wall so that the decay modes involving just the left-chiral fermions are not sufficiently suppressed. Furthermore, since the off-diagonal electroweak Yukawa constants are now non-zero, operators such $\overline{u_R}(\mu_R)\phi^*_c$ and  $\overline{s_R}(u_R)^c\phi_c$ are present in the action. This means we also have to account for the decay modes $p\rightarrow{}\mu^{+}\pi^{0}$, $p\rightarrow{}e^{+}K^0$ and $p\rightarrow{}\mu^{+}K^0$, which have partial lifetime lower bounds of $6.6\times{}10^{33}$ years \cite{superkprotondecaylimit2009}, $1.5\times{}10^{32}$ years \cite{pdgquark} and $1.2\times{}10^{32}$ years \cite{pdgquark}. After computing the overlaps to find the interaction strenghts in the weak eigenbasis, transforming to the mass eigenbasis and then performing similar analyses for each of the decay modes, we find that even the most constraining decay mode involving just right-chiral fermions, that of $\rightarrow{}\mu^{+}\pi^{0}$ sets a lower bound on the colored Higgs mass of $3.2\times{}10^9$ GeV. In fact both the modes involving the antimuon set higher bounds then those producing positrons since the coupling for the $\overline{u_R}(\mu_R)\phi^*_c$ vertex is a few orders of magnitude higher than that for the $\overline{u_R}(e_R)\phi^*_c$ vertex, due to right handed muon being closer on average to the wall. The decay modes involving left-chiral fermions are still barely suppressed, with the decay modes involving just left-chiral fermions for both $p\rightarrow{}e^{+}\pi^{0}$ and $p\rightarrow{}\mu^{+}\pi^{0}$ setting lower bounds on $m_c$ of order $10^{15}$ GeV. Obviously, this situation would change if we used the Higgs parameter choice of Eq. \ref{eq:thirdhiggschoice} for which the colored Higgs is well displaced from all fermions and the domain wall.

\subsection{Lepton mixing} 

  It appears that generating near tribimaximal mixing in the lepton sector is incompatible with the results for the fermion mass hierarchy problem, for both Dirac and Majorana neutrinos. 

 As we have seen, solutions to the mass hierarchy problem typically involve shifting the lepton doublets to different locations away from the domain wall and the electroweak Higgs. This means that for such solutions, assuming the domain wall couplings for the right handed neutrinos are roughly equal, the neutrino mass matrix will take the form,
\begin{equation}
M_{\nu} \sim{} \begin{pmatrix}
a_1 & b_1\epsilon{} & c_1\epsilon{}^2 \\
a_2 & b_2\epsilon{} & c_2\epsilon{}^2 \\
a_3 & b_3\epsilon{} & c_3\epsilon{}^2 \\ 
\end{pmatrix},
\end{equation}
where $\epsilon{}\ll{}1$ and the constants $a_i$, $b_i$ and $c_i$ are all taken to be roughly the same order of magnitude. Then the mass matrix squared will take the form
\begin{equation}
M_{\nu}^{\dagger}M_{\nu} \sim \begin{pmatrix}
|\mathbf{a}|^2 & \mathbf{a}.\mathbf{b}\epsilon & \mathbf{a}.\mathbf{c}\epsilon^2 \\
\mathbf{a}.\mathbf{b}\epsilon & |\mathbf{b}|^2\epsilon^2 & \mathbf{b}.\mathbf{c}\epsilon^3 \\
\mathbf{a}.\mathbf{c}\epsilon^2 & \mathbf{b}.\mathbf{c}\epsilon^3 & |\mathbf{c}|^2\epsilon^4 \\
\end{pmatrix}.
\end{equation}

 $M_{\nu}^{\dagger}M_{\nu}$ is clearly hierarchical, and thus the neutrino sector cannot generate two large mixing angles. From the electron sector, we know that both the lepton doublets and the right handed electrons are placed away from the electroweak Higgs doublet, and thus their overlaps decrease rapidly with the splittings, inducing hierarchical electron mass matrices. Hence, we cannot generate large mixing angles in the electron sector either, with the generic type of solution for the mass hierarchy, and thus tribimaximal mixing cannot be produced in the lepton sector for the case of a Dirac neutrino. 

 The utilization of the seesaw mechanism also fails to produce tribimaximal mixing. Since all the right handed neutrinos are all localized at the same place, all the overlap integrals which contribute to the Majorana mass matrix are of the same order of magnitude unless there is a substantial hierarchy amongst their domain wall parameters. Hence, for most solutions of interest, the right handed Majorana mass matrix assumes a non-hierarchal form. However, the neutrino Dirac mass matrices will maintain their hierarchical form since the lepton doublets will still be separated, and therefore the effective left handed neutrino Majorana mass matrix, $M_L \sim -m_D^TM_Rm_D$ is rendered hierarchical, and thus small lepton mixing angles for a Majorana neutrino will result. 

 There are several approaches one could take to the problem of lepton mixing in this model. The most obvious is the inclusion of a discrete flavor symmetry like $A_4$ or its double cover $T'$. This has in fact been employed successfully in $RS1$ \cite{rs1plusa4csaki} and orbifold models \cite{feruglioorbifolda4}. Such an inclusion is beyond the scope of this paper. 

 It was also not surprising that a solution was not found in the Dirac case, since the initial assumption that the couplings to $\eta$ and $\chi$ commuted cut the number of background Yukawa couplings to 15, and the assumption of universal electroweak Yukawa couplings cut the number of free parameters in the electroweak Yukawa sector to 1, giving 16 free parameters in total which determine the masses. If we want to generate everything except CP violation from non-hierarchical electroweak Yukawa couplings then, we need to generate the correct mass ratios for the charged fermions, quark and lepton mixing angles, the correct $\Delta{}m^2_{12}$ and $\Delta{}m^2_{23}$ and an acceptable neutrino mass scale, which amounts to 18 constraints. Thus, we were never guaranteed such a solution. Letting go of the initial assumption that the background couplings commute allows us to introduce mixing angles and CP phases from that sector. In practice, solving the relevant equations for the fermion profiles with non-commuting $h_{n\eta}$ and $h_{n\chi}$, is difficult and must be solved numerically; such an analysis will be deferred to a later paper.

\section{Conclusion} 
\label{sec:conclusion}

 As we have seen, the utilization of extra dimensions is particularly useful in explaining the fermion mass hierarchy problem. This is particularly evident in the analysis without quark and lepton mixing, where we were able to show that the mass hierarchy which spans at least 14 orders of magnitude could be generated from a set of domain wall Yukawa parameters which have a spread of roughly an order of magnitude. Furthermore, this spread could be reduced even further by making the Higgs profile more localized. As an added bonus in this analysis, by choosing parameters such that the colored Higgs was well displaced from the domain wall and the electroweak Higgs, the doublet-triplet splitting problem was solved and proton decay suppressed to such an extent that the colored Higgs mass no longer had to be fine tuned.

 Generation of quark mixing from the overlaps after initially assuming non-hierarchal $SU(5)$ electroweak Yukawa coupling constants also looks promising, and we successfully generated the Cabibbo angle and the fermion mass hierarchy for the case with two generations and Majorana neutrinos. Generically, small quark mixing angles and fermion mass hierarchies naturally arise from hierarchical mass matrices, although a more thorough numerical analysis will have to be done to find solutions for the full CKM matrix. 

 We have given some arguments as to why the problem of tribimaximal lepton mixing problem cannot be solved simultaneously with the quark mixing and fermion mass hierarchy problems in this braneworld model. Typically with solutions to the latter two problems, the lepton doublets are spread out away from the Higgs profile, rendering the electron and Dirac neutrino mass matrices hierarchical, leading to small mixing angles. We believe this may be amended with the addition of a discrete flavor symmetry like $A_4$, or by dropping the assumption that the $\eta$ and $\chi$ couplings commute.

 The addition of a flavor symmetry to the model as well as a more thorough analysis of the parameter space will be treated in later papers. We also must make a number of other additions to the analysis of this paper, most notably the inclusion of gravity and an analysis of the renormalization group evolution of the mass parameters.

 \subsection*{Acknowledgments}
 
 This work was supported by the Australian Research Council and the Commonwealth of Australia.  RRV would like to thank Andr\'{e} de Gouv\^{e}a and Ferruccio Feruglio for useful discussions.  Part of this work was performed at the Aspen Center for Physics.

\newpage

\bibliographystyle{ieeetr}
\bibliography{bibliography2.bib}

\end{document}